\documentclass[aps,12pt,showpacs,preprint,groupedaddress,floatfix]{revtex4}
\usepackage{graphicx}
\usepackage{dcolumn}
\usepackage{bm}
\usepackage{amssymb}

\newcommand{\bra}[1]{\left\langle #1 \right|}
\newcommand{\ket}[1]{\left| #1 \right\rangle}

\begin{document}

\title{Beyond the relativistic mean-field approximation (II):
configuration mixing of mean-field wave functions projected
on angular momentum and particle number}
\author{T. Nik\v si\' c}
\author{D. Vretenar}
\affiliation{Physics Department, Faculty of Science, University of Zagreb,
Croatia, and \\
Physik-Department der Technischen Universit\"at M\"unchen, D-85748 Garching,
Germany}
\author{P. Ring}
\affiliation{Physik-Department der Technischen Universit\"at M\"unchen,
D-85748 Garching,
Germany}
\date{\today}

\begin{abstract}
The framework of relativistic self-consistent mean-field
models is extended to include correlations related to the
restoration of broken symmetries and to fluctuations of
collective variables. The generator coordinate method is
used to perform configuration mixing of angular-momentum and
particle-number projected relativistic wave functions.
The geometry is restricted to axially symmetric shapes, and
the intrinsic wave functions are generated from the solutions of
the relativistic mean-field + Lipkin-Nogami
BCS equations, with a constraint on the mass quadrupole moment.
The model employs a relativistic point-coupling (contact)
nucleon-nucleon effective interaction in the particle-hole
channel, and a density-independent $\delta$-interaction in the
pairing channel. Illustrative calculations are performed
for $^{24}$Mg, $^{32}$S and $^{36}$Ar, and compared with results
obtained employing the model developed in the first part of this
work, i.e. without particle-number projection, as well as with
the corresponding non-relativistic models based on Skyrme and Gogny
effective interactions.
\end{abstract}

\maketitle

\section{\label{secI}Introduction}

In the first part of this work~\cite{NVR.06} we have extended the
theoretical framework of relativistic self-consistent mean-field
models to include correlations related to the restoration of broken
symmetries and to fluctuations of collective coordinates. In the
specific model which has been developed in~\cite{NVR.06}, the
generator coordinate method (GCM) is employed to perform
configuration mixing calculations of angular momentum projected wave
functions, calculated in a relativistic point-coupling model. The
geometry is restricted to axially symmetric shapes, and the mass
quadrupole moment is used as the generating coordinate. The
intrinsic wave functions are generated from the solutions of the
constrained relativistic mean-field + BCS equations in an axially
deformed oscillator basis. In order to test the implementation of
the GCM and angular momentum projection, a number of illustrative
calculations were performed for the nuclei $^{194}$Hg and $^{32}$Mg,
in comparison with results obtained in non-relativistic models based
on Skyrme and Gogny effective interactions.

In this work we develop the model further by including the restoration
of particle number in the wave functions of GCM states,
i.e. we restore a symmetry which is broken on the
mean-field level by the treatment of pairing correlations either
in the BCS approximation, or in the Hartree-Fock-Bogoliubov (HFB) framework.
We perform a GCM configuration mixing of angular-momentum and
particle-number projected relativistic wave functions. Projection on
particle number is crucial whenever the number of correlated pairs
becomes small and the density of levels close to the Fermi energy is
low, a situation typical for the description of phenomena related
to the evolution of shell structure \cite{BHR.03,VALR.05}:
reduction of spherical shell gaps and modifications of magic numbers
in nuclei far from stability, occurrence of islands of inversion
and coexistence of shapes with different deformations, moments of
inertia of superdeformed bands, etc. We thus plan to build a
self-consistent relativistic mean-field model in which rotational
symmetry and particle-number are restored, and fluctuations of the
quadrupole deformation are explicitly taken into account. Such a
model can be applied in a quantitative description of shell
evolution, and particularly in the treatment of shape coexistence
phenomena in nuclei with soft potential energy surfaces.

In Sec.~\ref{secII} we outline the relativistic point-coupling
model and the Lipkin-Nogami approximate particle number projection,
which are used to generate the intrinsic mean-field
wave functions with axial symmetry, and introduce the formalism of
configuration mixing of angular-momentum and particle-number projected
wave functions. In Sec.~\ref{secIII} the model is
tested in a detailed analysis of the spectra of $^{24}$Mg,
$^{32}$S and $^{36}$Ar. In order to illustrate the effects of
particle-number projection, we compare the results with those
obtained employing the model developed in the first part of this
work ~\cite{NVR.06}, in which the intrinsic wave functions are
generated from the solutions of the constrained relativistic
mean-field + BCS equations, without particle-number projection.
The results are also discussed in comparison with the
corresponding nonrelativistic GCM models based on Skyrme and
Gogny effective interaction. A brief summary and an outlook for
future studies are included in Sec.~\ref{secIV}.

\section{\label{secII}Theoretical framework}
\subsection{\label{subIIa}Implementation of the Lipkin-Nogami pairing scheme}

In the model that we have developed in Ref.~\cite{NVR.06}, the
intrinsic wave functions are generated from
constrained self-consistent solutions of the
relativistic mean-field (RMF) equations for the
point-coupling (PC) Lagrangian of Ref.~\cite{BMM.02}. Only basic
features of the RMF-PC model are outlined in \cite{NVR.06},
and we refer the reader to
\cite{BMM.02} and references therein, for a complete discussion of
the framework of relativistic point-coupling nuclear models. The
specific choice of the PC Lagrangian \cite{NVR.06,BMM.02} defines
the mean-field energy of a nuclear system
\begin{eqnarray}
{{E}}_{RMF} &=& \int d{\bm r }~{\mathcal{E}_{RMF}}(\bm{r}) \nonumber \\ &=&
\sum_k{\int d\bm{r}~v_k^2~{\bar{\psi}_k (\bm{r}) \left( -i\bm{\gamma}
 \bm{\nabla} + m\right )\psi_k(\bm{r})}} \nonumber \\
 &+& \int d{\bm r }~{\left (\frac{\alpha_S}{2}\rho_S^2+\frac{\beta_S}{3}\rho_S^3 +
  \frac{\gamma_S}{4}\rho_S^4+\frac{\delta_S}{2}\rho_S\triangle \rho_S
 + \frac{\alpha_V}{2}j_\mu j^\mu + \frac{\gamma_V}{4}(j_\mu j^\mu)^2 +
       \frac{\delta_V}{2}j_\mu\triangle j^\mu \right.} \nonumber \\
 &+& \left .
  \frac{\alpha_{TV}}{2}j^{\mu}_{TV}(j_{TV})_\mu+\frac{\delta_{TV}}{2}
    j^\mu_{TV}\triangle  (j_{TV})_{\mu}
 + \frac{\alpha_{TS}}{2}\rho_{TS}^2+\frac{\delta_{TS}}{2}\rho_{TS}\triangle
      \rho_{TS} +\frac{e}{2}\rho_p A^0
 \right) ,
\label{EMF}
\end{eqnarray}
where $\psi$ denotes the Dirac spinor field of a nucleon, and the
the local isoscalar and isovector densities and currents
\begin{eqnarray}
\label{dens_1}
\rho_{S}({\bm r}) &=&\sum_k v_k^2 ~\bar{\psi}_{k}({\bm r})
             \psi _{k}({\bm r})~,  \\
\label{dens_2}
\rho_{TS}({\bm r}) &=&\sum_k v_k^2 ~
      \bar{\psi}_{k}({\bm r})\tau_3\psi _{k}^{{}}({\bm r})~,  \\
\label{dens_3}
j^{\mu}({\bm r}) &=&\sum_k v_k^2 ~\bar{\psi}_{k}({\bm r})
        \gamma^\mu\psi _{k}^{{}}({\bm r})~,  \\
\label{dens_4}
j^{\mu}_{TV}({\bm r}) &=&\sum_k v_k^2 ~\bar{\psi}_{k}({\bm r})
     \gamma^\mu \tau_3 \psi _{k}^{{}}({\bm r})~,
\end{eqnarray}
are calculated in the {\it no-sea} approximation: the summation
runs over all occupied states in the Fermi sea, i.e. only
occupied single-nucleon states with positive energy explicitly
contribute to the nucleon self-energies.
$v_k^2$ denotes the occupation factors of single-nucleon states.
In Eq.~(\ref{EMF}) $\rho_p$ is the proton density,
and $A^0$ denotes the Coulomb potential.

In addition to the self-consistent mean-field potential, for
open-shell nuclei pairing correlations have to be included in the
energy functional. In this work we do not consider nuclear systems
very far from the valley of $\beta$-stability, and therefore a good
approximation for the treatment of pairing correlations is provided
by the BCS formalism. Following the prescription from
Ref.~\cite{BMM.02}, we use a $\delta$-interaction in the pairing
channel, supplemented with a smooth cut-off determined by the Fermi
function of single-particle energies $\epsilon_k$ \cite{KBF.90}:
\begin{equation}
f_k^2=\frac{1}{1+e^{(\epsilon_k-\lambda_\tau-\Delta E_\tau)/\mu_\tau}}\;,
\label{fk2}
\end{equation}
where $\lambda_\tau$ is the Fermi energy for neutrons ($\tau=n$) or
protons ($\tau=p$). $f^2_k$ is used in the evaluation of the pairing
density
\begin{equation}
\kappa_\tau(\bm{r})=-2\sum_{k>0}f_ku_kv_k\psi_k(\bm{r})^\dagger\psi_k(\bm{r})\;,
\label{pten}
\end{equation}
where the summation runs for $\tau=n(p)$ over neutron (proton)
single-particle states. The cut-off parameters $\Delta E_q$ and
$\mu_q = \Delta E_q/10$ are adjusted to the density of
single-particle levels in the vicinity of the Fermi energy. In
particular, the sum of the cut-off weights approximately includes
one additional shell of single-particle states above the Fermi level
\begin{equation}
\sum_{k>0}2f_k = N^{}_\tau+1.65N_\tau^{2/3}\;,
\end{equation}
where $N_\tau$ denotes number of neutrons (protons) in a specific
nucleus. The pairing contribution to the total energy is given by
\begin{equation}
{E}_{pair}^\tau = \int{\mathcal{E}_{pair}^\tau(\bm{r})d\bm{r}}=
   \frac{V_\tau}{4}\int{\kappa_\tau^*(\bm{r})\kappa_\tau(\bm{r}) d\bm{r}}\;,
\label{epair}
\end{equation}
where $V_{n(p)}$ denotes the strength parameter of the pairing
interaction for neutrons (protons).
Finally, the expression for the total energy reads
\begin{equation}
E_{tot} =\int{\left[\mathcal{E}_{RMF}(\bm{r})+\mathcal{E}_{pair}^p(\bm{r})
         +\mathcal{E}_{pair}^n(\bm{r})\right]d\bm{r} }\;,
\label{etot}
\end{equation}
and the center-of-mass correction is included by adding the expectation
value
\begin{equation}
 E_{cm} = -\frac{\langle \hat{\bm{P}}_{cm}^2 \rangle}{2mA}\;,
\end{equation}
to the total energy, where $\bm{P}_{cm}$ denotes the total momentum of the
nucleus.

The principal disadvantage of describing pairing correlations in the
BCS approximation is that the resulting wave function is not an
eigenstate of the particle number operator. More precisely, the BCS
ground state contains admixtures of particle-number eigenstates with
a relative spread of order $1/\sqrt{N}$, where $N$ denotes the
average number of valence particles. The ideal solution, of course,
is to perform particle number projection from the BCS state before
variation. This procedure is technically rather complicated and very
much time consuming, and therefore it is usual to employ the
Lipkin-Nogami (LN) approximation to the exact particle number
projection~\cite{Lip.60,Nog.64,FO.97}. In this work the LN method is
implemented in terms of local density functionals of the effective
interaction, as developed in
Refs.~\cite{RNB.96,VER.96,VER.97,BRR.00}.

The Lipkin-Nogami equations are obtained from the variation of the functional
\begin{equation}
\mathcal{K} = \mathcal{E}_{tot}
       -\sum_{\tau=n,p}{\lambda_{1,\tau}\langle\hat{N}^{}_\tau\rangle+
        \lambda_{2,\tau}\langle \hat{N}_\tau^2\rangle}\;,
\label{LNfunctional}
\end{equation}
with respect to the single-particle states $\bar{\psi}_k$ and the occupation
amplitudes $v_k$. $\mathcal{E}_{tot}$ is the total energy functional
of Eq.~(\ref{etot}).
The resulting expression for the occupation probabilities
can be cast into the standard BCS formula
\begin{equation}
v_k^2=\frac{1}{2}\left[ 1-\frac{\epsilon^\prime_k-\lambda_\tau}
    {\sqrt{(\epsilon^\prime_k-\lambda_\tau)^2+f_k^2\Delta_k^2}}\right]\;,
\label{LNv2}
\end{equation}
where $\epsilon^\prime_k=\epsilon_k+4\lambda_{2,\tau}v_k^2$ denotes
the renormalized single-particle energy, and
$\lambda_\tau=\lambda_{1,\tau}+4\lambda_{2,\tau}(N_\tau+1)$ is the
generalized Fermi energy. $\lambda_{1,\tau}$ is determined by a
particle number subsidiary condition such that the expectation value
of the particle number operator equals the given number of nucleons.
The state-dependent single-particle gaps are defined as the matrix
elements
\begin{equation}
\Delta_k =
\int{d\bm{r}\psi_k^\dagger(\bm{r})\Delta_\tau(\bm{r})\psi_k(\bm{r})}\;,
\label{gap}
\end{equation}
of the local pair potential
\begin{equation}
\Delta_\tau(\bm{r})=\frac{1}{2}V_\tau\kappa_\tau(\bm{r})\quad \quad
(\tau\equiv n,p)\;. \label{pairpot}
\end{equation}
While the quantities $\lambda_{1,\tau}$ represent the Lagrange
multipliers used to constrain the average particle numbers, the
value of the parameters $\lambda_{2,\tau}$ are determined from
\begin{equation}
\lambda_{2,\tau} = \frac{\langle
\hat{H}\Delta\hat{N}_{2,\tau}^2\rangle}
    {\langle \hat{N}_\tau^2\Delta\hat{N}_{2,\tau}^2\rangle}\;,
\label{lam2}
\end{equation}
where $\hat{N}_{2,\tau}$ is the term of the particle number operator
which projects onto two-quasiparticle states
\begin{equation}
\hat{N}_{2,\tau}=2\sum_{k>0}{u_k
v_k(\alpha_k^\dagger\alpha_{\bar{k}}^\dagger+\alpha^{}_{\bar{k}}\alpha^{}_k)}\;,
\label{n2}
\end{equation}
and
$\Delta\hat{N}_{2,\tau}^2=\hat{N}_{2,\tau}^2-\langle\hat{N}_{2,\tau}^2\rangle$
denotes its variance. The evaluation of the parameters
$\lambda_{2,\tau}$ is described in the Appendix.

After approximate particle-number projection the total binding energy reads
\begin{equation}
{E}_{LN} =
{E}_{tot}-\sum_{\tau=n,p}\lambda_{2,\tau}\langle(\Delta\hat{N}_{2,\tau})^2\rangle\;.
\label{etotLN}
\end{equation}
The strength of the pairing interaction can be determined by comparing the
average pairing gaps with empirical gaps obtained from nuclear masses.
The average gap is given by the summation over occupied states with
either the occupation probability
$v_k^2$ as the weighting factor \cite{DFT.84}
\begin{equation}
\langle v^2\Delta\rangle_\tau = \frac{\sum_{k>0}
f_kv_k^2\Delta_k}{\sum_{k>0}
      f_kv_k^2}\;,
\label{v2gap}
\end{equation}
or the factor $u_kv_k$~\cite{BRR.00}
\begin{equation}
\langle uv\Delta\rangle_\tau = \frac{\sum_{k>0}
f_ku_kv_k\Delta_k}{\sum_{k>0}
      f_ku_kv_k}\;.
\label{uvgap}
\end{equation}
The corresponding expressions for the (approximately)
particle-number projected average pairing gaps
read~\cite{CDH.96,BRR.00}:
\begin{equation}
\langle v^2\Delta\rangle_\tau^{LN} = \langle v^2\Delta\rangle_\tau
+\lambda_{2,\tau}\;, \label{v2gapPR}
\end{equation}
\begin{equation}
\langle uv\Delta\rangle_\tau^{LN} = \langle
uv\Delta\rangle_\tau+\lambda_{2,\tau}\;. \label{uvgapPR}
\end{equation}
The local densities and currents which define the energy functional
refer to the intrinsic state, and are therefore computed from the
Eqs. (\ref{dens_1}) -- (\ref{dens_4}). On the other hand, the
densities used to evaluate physical observables, such as the mass
quadrupole moment, must correspond to the (approximately)
particle-number projected state. The LN density is simply computed
by replacing the occupation probabilities $v_k^2$, with the LN
occupation coefficients~\cite{BHB.89,QRM.90}
\begin{equation}
w_k = v_k^2 + \frac{u_k^2v_k^2\overline{u^2v^2}\left[ (v_k^2-u_k^2)
    \overline{u^2v^2} -\overline{u^2v^2(v^2-u^2)}\right]}
 {\overline{u^2v^2}\;\overline{u^2v^2(v^2-u^2)^2}-[\overline{u^2v^2(v^2-u^2)}]^2
  +2\overline{u^2v^2}[(\overline{u^2v^2})^2-\overline{u^4v^4}]}\;,
\end{equation}
where $\overline{x}$ denotes half of the trace
\begin{equation}
\overline{x}=\sum_{k>0}x_k\;.
\end{equation}
We note that LN-corrected quadrupole moments will be used in the
constrained self-consistent relativistic mean-field
calculations (see Eq.~\ref{constr}).

In this work we only consider even-even nuclei that can be described
by axially symmetric shapes. In addition to axial symmetry and
parity, symmetry with respect to the operator $e^{-i\pi \hat{J}_y}$,
and time-reversal invariance are imposed as self-consistent
symmetries. The single-nucleon Dirac eigenvalue equation is solved
by expanding the large and small components of the nucleon spinor
$\psi_i$ in terms of eigenfunctions of an axially symmetric harmonic
oscillator potential (see Ref.~\cite{GRT.90} for details).
\subsection{\label{subIIb}Configuration mixing of mean-field solutions projected
           on angular momentum and particle number}

Correlation effects related to the restoration of broken symmetries and
to fluctuations of collective coordinates can be taken into account by
performing configuration mixing calculations of projected states. The
generator coordinate method (GCM), which uses a set of mean-field states
$\ket{\phi(q)}$ that depend on a collective coordinate $q$, provides a
very efficient procedure for the construction of the trial wave
function~\cite{RS.80}:
\begin{equation}
\ket{\Psi_\alpha} = \sum_j{f_\alpha(q_j)\ket{\phi (q_j)}}\;.
\label{GCM-state}
\end{equation}
In this work the basis states $\ket{\phi (q)}$ are Slater
determinants of single-nucleon states generated by solving the
constrained relativistic mean-field + LNBCS equations, with
the quadrupole moment as the generating coordinate $q$.
For an axially deformed nucleus the map of the energy surface
as a function of deformation is obtained by imposing a
constraint on the mass quadrupole moment. The method of quadratic
constraint uses an unrestricted variation of the function
\begin{equation}
\langle \mathcal{K}\rangle~+~\frac{C}{2}\left( \langle\hat{Q} \rangle - ~q \right)^2 \;,
\label{constr}
\end{equation}
where $\langle\mathcal{K}\rangle$ is the energy functional
of Eq.~(\ref{LNfunctional}),
 $\langle\hat{Q}\rangle$
denotes the expectation value of the mass quadrupole operator, $q$ is the
deformation parameter, and $C$ is the stiffness constant.

The axially deformed mean-field breaks rotational symmetry, and the particle
number is only approximately restored with the Lipkin-Nogami procedure,
i.e. the basis states $\ket{\phi(q)}$ are not eigenstates of
the total angular momentum and particle number operators. Therefore,
in order to be able to compare model predictions with data, we must
construct states with good angular momentum and particle number, by
performing projections from the mean-field plus LNBCS solutions
\begin{equation}
\ket{\Psi_\alpha^{JM}} = \sum_{j,K}{f_\alpha^{JK}(q_j)\hat{P}_{MK}^J
     \hat{P}^Z \hat{P}^N \ket{\phi (q_j)}}~.
\label{PNPAMPGCM-state}
\end{equation}
The particle-number projection operators read
\begin{equation}
\hat{P}^N=\frac{1}{2\pi}\int_0^{2\pi}{d\varphi_n
e^{i(\hat{N}-N)\varphi_n}}~,\quad
\hat{P}^Z=\frac{1}{2\pi}\int_0^{2\pi}{d\varphi_p
e^{i(\hat{Z}-Z)\varphi_p}}~. \label{pnp_op}
\end{equation}
where $\hat{N}(\hat{Z})$ is the number operator for neutrons
(protons), and $N(Z)$ denotes the number of neutrons (protons).

The angular momentum projection operator is defined by
\begin{equation}
\hat{P}^J_{MK} = \frac{2J+1}{8\pi^2}\int{d\Omega D_{MK}^{J*}(\Omega )\hat{R}
   (\Omega )}\;,
\label{amp_op}
\end{equation}
where the integration is performed over the three Euler angles
$\alpha$, $\beta$, and $\gamma$. $D_{MK}^{J}(\Omega
)=e^{-iM\alpha}d^J_{MK}(\beta)e^{-iK\gamma}$ is the Wigner
function~\cite{VMK.88}, and $\hat{R}(\Omega
)=e^{-i\alpha\hat{J}_z}e^{-i\beta\hat{J}_y} e^{-i\gamma\hat{J}_z}$
is the rotation operator.

The weight functions $f_{\alpha}^{JK}(q_j)$ are determined by requiring that
the expectation value of the energy is stationary
with respect to an arbitrary variation $\delta f_{\alpha}^{JK}$:
\begin{equation}
 \delta E^{J} =
 \delta \frac{\bra{\Psi_\alpha^{JM}} \hat{H} \ket{\Psi_\alpha^{JM}}}
            {\bra{\Psi_\alpha^{JM}}\Psi_\alpha^{JM}\rangle} = 0 \; .
\label{variational}
\end{equation}
This leads to the Hill-Wheeler equation~\cite{HW.53}:
\begin{equation}
\sum_{j,K}f_{\alpha}^{JK}(q_j)
  \left( \left\langle\phi(q_i) \right|\hat{H}\hat{P}_{MK}^J\hat{P}^N\hat{P}^Z\left|
  \phi(q_j)\right\rangle - E^J_\alpha
  \left\langle\phi(q_i) \right|\hat{P}_{MK}^J\hat{P}^N\hat{P}^Z
   \left|\phi(q_j)\right\rangle \right) = 0\;.
\label{HWEQ}
\end{equation}
The restriction to axially symmetric configurations
($\hat{J}_z\ket{\phi(q)} = 0$) simplifies the problem considerably, because
in this case the integrals over the Euler angles $\alpha$ and $\gamma$ can be
performed analytically. In addition, the symmetry with respect to the
operator $e^{-i\pi \hat{J}_y}$ reduces the integration interval over the Euler
angle $\beta$ from $[0,\pi ]$ to $[0,\pi /2]$. For an arbitrary
multipole operator $\hat{Q}_{\lambda \mu}$ one thus finds
\begin{eqnarray}
&&\bra{\phi(q_i)}\hat{Q}_{\lambda \mu} \hat{P}_{MK}^J\hat{P}^N\hat{P}^Z\ket{\phi(q_j)}
=(2J+1)\frac{1+(-1)^J}{2}\delta_{M-\mu}\delta_{K0}\times \frac{1}{(2\pi)^2}\nonumber \\
 && \int_0^{2\pi}d\varphi_n \int_0^{2\pi}d\varphi_p
  \int_0^{\pi/2}\sin\beta d_{-\mu 0}^{J*}(\beta) \bra{\phi(q_i)}
  \hat{Q}_{\lambda \mu}e^{-i\beta\hat{J}_y}
  e^{i(\hat{N}-N)\varphi_n}e^{i(\hat{Z}-Z)\varphi_p}
  \ket{\phi(q_j)}d\beta\;.
\label{matel}
\end{eqnarray}
We note that this expression is defined only for even values
of the angular momentum $J$. The norm overlap kernel
\begin{eqnarray}
\mathcal{N}^J(q_i,q_j) &=& \bra{\phi(q_i)} \hat{P}_{MK}^J\hat{P}^N\hat{P}^Z
        \ket{\phi(q_j)} = \nonumber \\ &&
(2J+1)\frac{1+(-1)^J}{2}\delta_{M0}\delta_{K0}\frac{1}{(2\pi)^2}
    \int_0^{2\pi}d\varphi_n\int_0^{2\pi}d\varphi_p  \times \nonumber \\ &&
  \int_0^{\pi/2}{\sin{\beta}d_{00}^{J*}(\beta) \bra{\phi(q_i)}
  e^{-i\beta\hat{J}_y}e^{i(\hat{N}-N)\varphi_n}e^{i(\hat{Z}-Z)\varphi_p}
  \ket{\phi(q_j)}d\beta}\;,
\label{normker}
\end{eqnarray}
can be evaluated by using the fact, that $\ket{\phi(q_j)}$ is a
product of a neutron- and a proton Slater determinant and the
generalized Wick theorem~\cite{OY.66,BB.69,BDF.90,VHB.00}:
\begin{eqnarray}
n(q_i,q_j;\beta,\varphi_\tau) &\equiv& \bra{\phi(q_i)}
e^{-i\beta\hat{J}_y}
    e^{i\hat{N}_\tau\varphi_\tau}\ket{\phi(q_j)}
    = \pm\sqrt{det~\mathcal{N}_{ab}(q_i,q_j;\beta,\varphi_\tau)}\;,
\label{norm}
\end{eqnarray}
for $\tau=n,p$. The overlap matrix is defined as
\begin{equation}
\mathcal{N}_{ab}(q_i,q_j;\beta,\varphi_\tau) =
            u_a(q_i)R_{ab}(q_i,q_j;\beta)u_b(q_j) +
            v_a(q_i)R_{ab}(q_i,q_j;\beta)v_b(q_j)e^{2i\varphi_\tau} \;,
\label{N_mat}
\end{equation}
where $u$ and $v$ denote the BCS occupation probabilities, and the
elements of the matrix $R$ read
\begin{equation}
R_{ab}(q_i,q_j;\beta) = \int{\psi_a^\dagger(\bm{r};q_i)e^{-i\beta\hat{J}_y}
                     \psi_b(\bm{r};q_j)d\bm{r}}\;.
\label{R_mat}
\end{equation}
We note that the global phase of the overlap in Eq.~(\ref{norm}) is
determined by using the procedure described in Ref.~\cite{VHB.00}.
The details of the evaluation of the matrix $R$ can be found in
Ref.~\cite{NVR.06}.

The Hamiltonian kernel
\begin{eqnarray}
\mathcal{H}^J(q_i,q_j) &=& \bra{\phi(q_i)}\hat{H} \hat{P}_{MK}^J\hat{P}^N\hat{P}^Z
      \ket{\phi(q_j)} =\nonumber \\ &&
(2J+1)\frac{1+(-1)^J}{2}\delta_{M0}\delta_{K0} \frac{1}{(2\pi)^2}
  \int_0^{2\pi}d\varphi_n\int_0^{2\pi}d\varphi_p  \times \nonumber \\ &&
  \int_0^{\pi/2}\sin{\beta}d_{00}^{J*}(\beta)
 \bra{\phi(q_i)}\hat{H}
  e^{-i\beta\hat{J}_y}e^{i(\hat{N}-N)\varphi_n}e^{i(\hat{Z}-Z)\varphi_p}
   \ket{\phi(q_j)}d\beta\;,
\label{hamker}
\end{eqnarray}
can be calculated from the mean-field energy functional
Eq.~(\ref{EMF}), provided the modified
densities~\cite{OY.66,BB.69,BDF.90,VHB.00}
\begin{eqnarray}
\tau^\tau(\bm{r};q_i,q_j,\beta,\varphi_\tau) &=&
      \sum_{a,b}v_a(q_i)v_b(q_j)e^{2i\varphi_\tau}
      \mathcal{N}_{ba}^{-1}(q_i,q_j;\beta) \times\nonumber \\ & &
      \bar{\psi}_a(\bm{r};q_i)
      (-i\bm{\gamma}\bm{\nabla} + m)
      e^{-i\beta\hat{J}_y}\psi_b(\bm{r};q_j)\;, \\
\rho_{S}^\tau(\bm{r};q_i,q_j,\beta,\varphi_\tau) &=&
      \sum_{a,b}v_a(q_i)v_b(q_j)e^{2i\varphi_\tau}
      \mathcal{N}_{ba}^{-1}(q_i,q_j;\beta)
      \bar{\psi}_a(\bm{r};q_i)
      e^{-i\beta\hat{J}_y}\psi_b(\bm{r};q_j) \;, \\
j_\mu^\tau(\bm{r};q_i,q_j,\beta,\varphi_\tau) &=&
      \sum_{a,b}v_a(q_i)v_b(q_j)e^{2i\varphi_\tau}
      \mathcal{N}_{ba}^{-1}(q_i,q_j;\beta)
      \bar{\psi}_a(\bm{r};q_i)\gamma_\mu
      e^{-i\beta\hat{J}_y}\psi_b(\bm{r};q_j) \;,
\label{dens_gcm}
\end{eqnarray}
are used when evaluating the expression
\begin{eqnarray}
h(q_i,q_j;\beta,\varphi_n,\varphi_p)&\equiv&
  \bra{\phi(q_i)}\hat{H}
  e^{-i\beta\hat{J}_y}e^{i\hat{N}\varphi_n}e^{i\hat{Z}\varphi_p}\ket{\phi(q_j)}
  =\int{\mathcal{E}_{tot}(\bm{r};q_i,q_j,\beta,\varphi_n,\varphi_p)d\bm{r}}.
\end{eqnarray}
The computational task of evaluating the Hamiltonian and norm
overlap kernels can be reduced significantly if one realizes that
states with very small occupation probabilities give negligible
contributions to the kernels, and hence such states can be excluded
from the calculation~\cite{BDF.90,VHB.00}.

For an even number of particles, the integration interval in Eq.~(\ref{pnp_op})
can be reduced to $[0,\pi]$ using the symmetries of the integrand. 
Furthermore, the integrals can be discretized
by using the Fomenko's expression~\cite{Fom.70}
\begin{equation}
\hat{P}^N=\frac{1}{L}\sum_{n=1}^{L}e^{i(\hat{N}-N)\varphi_n},\qquad
     \varphi_n=\frac{\pi}{L}n~,
\label{fomenko}
\end{equation}
with $L$ points in the expansion. In order to avoid numerical
instabilities which might arise at $\varphi = \frac{\pi}{2}$ when
the occupation probability of a state is exactly $0.5$, an odd
number of points must be used in the Fomenko's
expansion~\cite{AER.01}. The Gauss-Legendre quadrature is used for
the integration over the Euler angle $\beta$. We have verified that
already $L=9$ points both for protons and neutrons in the Fomenko's
expansion, and $13$ points in the integral over $\beta$, produce
numerically stable results.

The Hill-Wheeler equation (\ref{HWEQ})
\begin{equation}
\sum_j{\mathcal{H}^J(q_i,q_j)f^J_\alpha(q_j)} = E^J_\alpha
        \sum_j{\mathcal{N}^J(q_i,q_j)f^J_\alpha(q_j)}\;,
\label{HWEQ2}
\end{equation}
presents a generalized eigenvalue problem. Thus the weight functions
$f^J_\alpha(q_i)$ are not orthogonal and cannot be interpreted as
collective wave functions for the variable $q$. The standard
procedure~\cite{Lat.76} is to re-express Eq. (\ref{HWEQ2}) in terms
of another set of functions,  $g^J_\alpha(q_i)$, defined by
\begin{equation}
g^J_\alpha(q_i) = \sum_j(\mathcal{N}^{J})^{1/2}(q_i,q_j) f^J_\alpha(q_j)\;.
\label{coll_wf}
\end{equation}
With this transformation the
Hill-Wheeler equation defines an ordinary eigenvalue problem
\begin{equation}
\sum_j{\tilde{\mathcal{H}}^J(q_i,q_j)g^J_\alpha(q_j)} =E_\alpha g_\alpha^J(q_i)\;,
\end{equation}
with
\begin{equation}
\tilde{\mathcal{H}}^J(q_i,q_j) = \sum_{k,l}(\mathcal{N}^{J})^{-1/2}(q_i,q_k)
  \mathcal{H}^J(q_k,q_l) (\mathcal{N}^{J})^{-1/2}(q_l,q_j) \;.
\end{equation}
The functions $g^J_\alpha(q_i)$ are orthonormal and play the role
of collective wave functions. For a more detailed description of
this particular implementation of the
Hill-Wheeler equation, we refer the reader to Ref.~\cite{NVR.06}.

For completeness we also include the expressions for physical
observables, such as transition probabilities and spectroscopic
quadrupole moments~\cite{RER.02b}. The reduced transition
probability for a transition between an initial state
$(J_i,\alpha_i)$, and a final state $(J_f,\alpha_f)$, reads
\begin{equation}
B(E2; J_i\alpha_i \to J_f\alpha_f) = \frac{e^2}{2J_i+1}\left| \sum_{q_f,q_i}{
f^{J_f*}_{\alpha_f}(q_f)\bra{J_fq_f}|\hat{Q}_2|\ket{J_iq_i}
f^{J_i}_{\alpha_i}(q_i)}\right|^2\;,
\label{BE2}
\end{equation}
and the spectroscopic quadrupole moment for a state $(J\alpha)$ is defined
\begin{equation}
Q^{spec}(J,\alpha)=e\sqrt{\frac{16\pi}{5}}
\left( \begin{array}{ccc}
J & 2 & J \\
J & 0 & -J \end{array} \right)
\sum_{q_i,q_j}{f^{J*}_{\alpha}(q_i)\bra{Jq_i}|\hat{Q}_2|\ket{Jq_j}
f^{J}_{\alpha}(q_j)}\;.
\label{Qspec}
\end{equation}
Since these quantities are calculated in full configuration space,
there is no need to introduce effective charges, hence $e$ denotes
the bare value of the proton charge. In order to evaluate transition
probabilities and spectroscopic quadrupole moments, we will
also need the reduced matrix element of the quadrupole operator
\begin{eqnarray}
\bra{J_fq_f}|\hat{Q}_2|\ket{J_iq_i}&=&(2J_i+1)(2J_f+1)\sum_\mu
\left( \begin{array}{ccc}
J_i & 2 & J_f \\
-\mu & \mu & 0 \end{array} \right) \frac{1}{(2\pi)^2}
   \int_0^{2\pi}d\phi_n\int_0^{2\pi}d\phi_p \times \nonumber \\ &&
\int^{\pi/2}_0 {\sin{\beta}d_{-\mu0}^{J_i*}(\beta)\bra{\phi(q_f)}\hat{Q}_{2\mu}
 e^{-i\beta\hat{J}_y}e^{i(\hat{N}-N)\phi_n}e^{i(\hat{Z}-Z)\phi_p}\ket{\phi(q_i)}d\beta}.
\label{Q2red}
\end{eqnarray}
\section{\label{secIII}Illustrative calculations}
\begin{figure}
\includegraphics[scale=0.6]{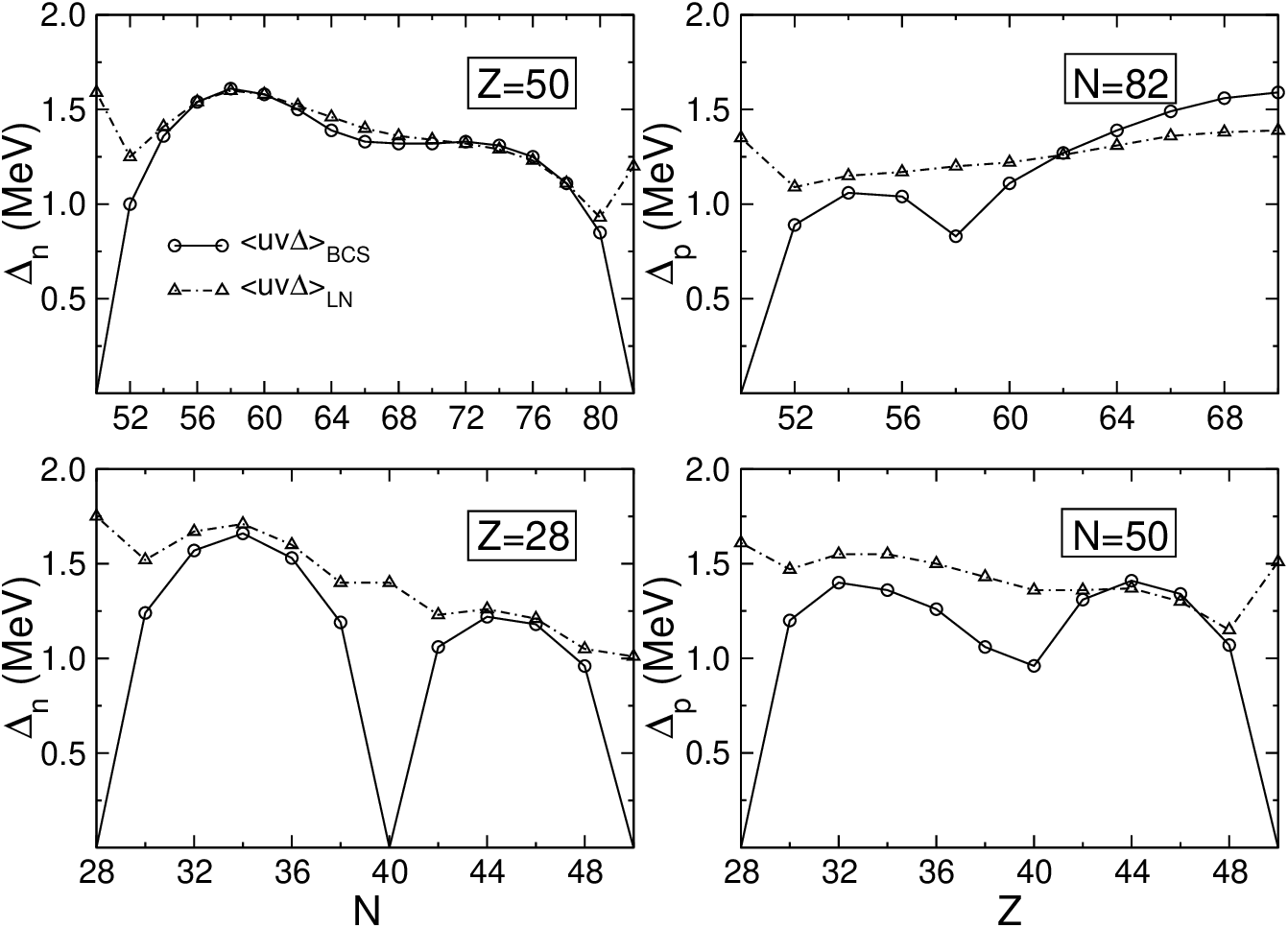}
\caption{Comparison of the average pairing gaps calculated with a
simple BCS, and the Lipkin-Nogami plus BCS approximations.
For the $Z=50$ and $Z=28$ isotopes the neutron pairing gaps are
shown, whereas for the $N=50$ and $N=82$ isotones the
proton average pairing gaps are compared.}
\label{figA}
\end{figure}

The intrinsic wave functions that will be used in configuration
mixing calculations are obtained as solutions of the self-consistent
RMF+LNBCS equations, subject to constraint on the mass quadrupole
moment. As in the first part of this analysis \cite{NVR.06}, we use
the relativistic point-coupling interaction PC-F1 \cite{BMM.02} in
the particle-hole channel, and a density-independent $\delta$-force
is the effective interaction in the particle-particle channel. The
parameters of the PC-F1 interaction and the pairing strength
constants $V_n$ and $V_p$ have been adjusted simultaneously to the
nuclear matter equation of state, and to ground-state observables
(binding energies, charge and diffraction radii, surface thickness
and pairing gaps) of spherical nuclei \cite{BMM.02}, with pairing
correlations treated in the BCS approximation. However, since the
present analysis includes the LN approximate particle number
projection, the pairing strength parameters have to be readjusted.
By comparing the projected average pairing gaps $\langle uv\Delta
\rangle^{(LN)}_\tau$ and the BCS pairing gaps $\langle uv\Delta
\rangle^{(BCS)}_\tau$, we find that the neutron pairing strength
should be reduced from $V_n = -308$ MeV to $V_n = -285$ MeV, and the
proton pairing strength from $V_p = -321$ MeV to $V_p = -260$ MeV.
The average pairing gaps for two isotopic and two isotonic chains
are shown in Fig.~\ref{figA}. We notice that by readjusting the
strength parameters $V_n$ and $V_p$, a good agreement between
projected average pairing gaps and the BCS average pairing gaps is
obtained except, of course, for magic numbers of neutrons or
protons, for which an unphysical collapse of pairing correlations is
found in the BCS approximation.

In order to illustrate the importance of including particle-number
projection in the description of specific structure effects, we will
compare the results of GCM calculations with those obtained
using the model that was developed in the first part of this
work \cite{NVR.06}, and in which the intrinsic wave functions are
generated from the solutions of the constrained relativistic
mean-field + BCS equations, without the Lipkin-Nogami
approximate particle-number projection. In that case the
correct mean-values of the nucleon numbers are restored by
modifying the Hill-Wheller equations to include linear constraints
on the number of protons and neutrons (see Eqs. (54) and (55) of
Ref.~\cite{NVR.06}). Therefore, in the following subsections AMP
will indicate that only angular momentum projection has been carried out
before GCM configuration mixing (i.e. the model of Ref.~\cite{NVR.06}
is used), and PN\&AMP will denote the results of GCM calculations which
include both the restoration of the particle number and rotational symmetry.

In this section illustrative configuration mixing calculations are
presented for $^{24}$Mg, $^{32}$S and $^{36}$Ar. We choose these
nuclei because the results can be directly compared with extensive
GCM studies performed using the nonrelativistic Skyrme and Gogny
effective interactions. In the analyses in which the nonrelativistic
zero-range Skyrme interaction was used \cite{VHB.00,BFH.03}, both
particle number and angular momentum projections were performed. The
simultaneous projection on particle number and angular momentum is
computationally much more demanding in the case of a finite-range
interaction, and therefore only angular momentum projection was
performed in the studies with the Gogny
force~\cite{RER.00,RER.02b,RER.04}. Both approaches are interesting
for the present analysis, because by comparing the results one can
deduce which effects can be attributed to particle number
projection, and which originate from the differences in the
properties of the effective interactions.

The constrained RMF equations are solved by expanding the Dirac
single-nucleon spinors in terms of eigenfunctions of an axially
symmetric harmonic oscillator potential. In order to keep the basis
closed under rotations, the two oscillator length parameters
$b_\perp$ and $b_z$ have always identical values~\cite{Rob.94,
ERS.93}. In addition, to avoid the completeness problem in
subsequent configuration mixing calculations, the same oscillator
length is used for all values of the quadrupole
deformations~\cite{ERS.93}. Since we consider relatively light
nuclei in this work, it is sufficient to use ten oscillator shells
in the expansion (see Ref.~\cite{NVR.06} for details).

\subsection{\label{subIIIa}$^{24}$Mg}
The low-energy spectrum of $^{24}$Mg displays a typical rotational
structure, with data on the ground-state band extending up to
angular momentum $I=8\hbar$~\cite{Endt.90,EBF.98}. The spectroscopic
quadrupole moment of the $2_1^+$ state: $-16.6(6)$ e
fm${^2}$~\cite{EBF.98}, indicates a large prolate deformation. The
$0_2^+$ state is found at high excitation energy ($6.432$ MeV), and
this means that shape-coexistence effects should only play a minor
role in the description of the ground-state band. Properties of
$^{24}$Mg have been studied with the GCM using the Skyrme
SLy4~\cite{VHB.00}, and the Gogny D1S~\cite{RER.02b} effective
interactions.

\begin{figure}
\includegraphics[scale=0.6]{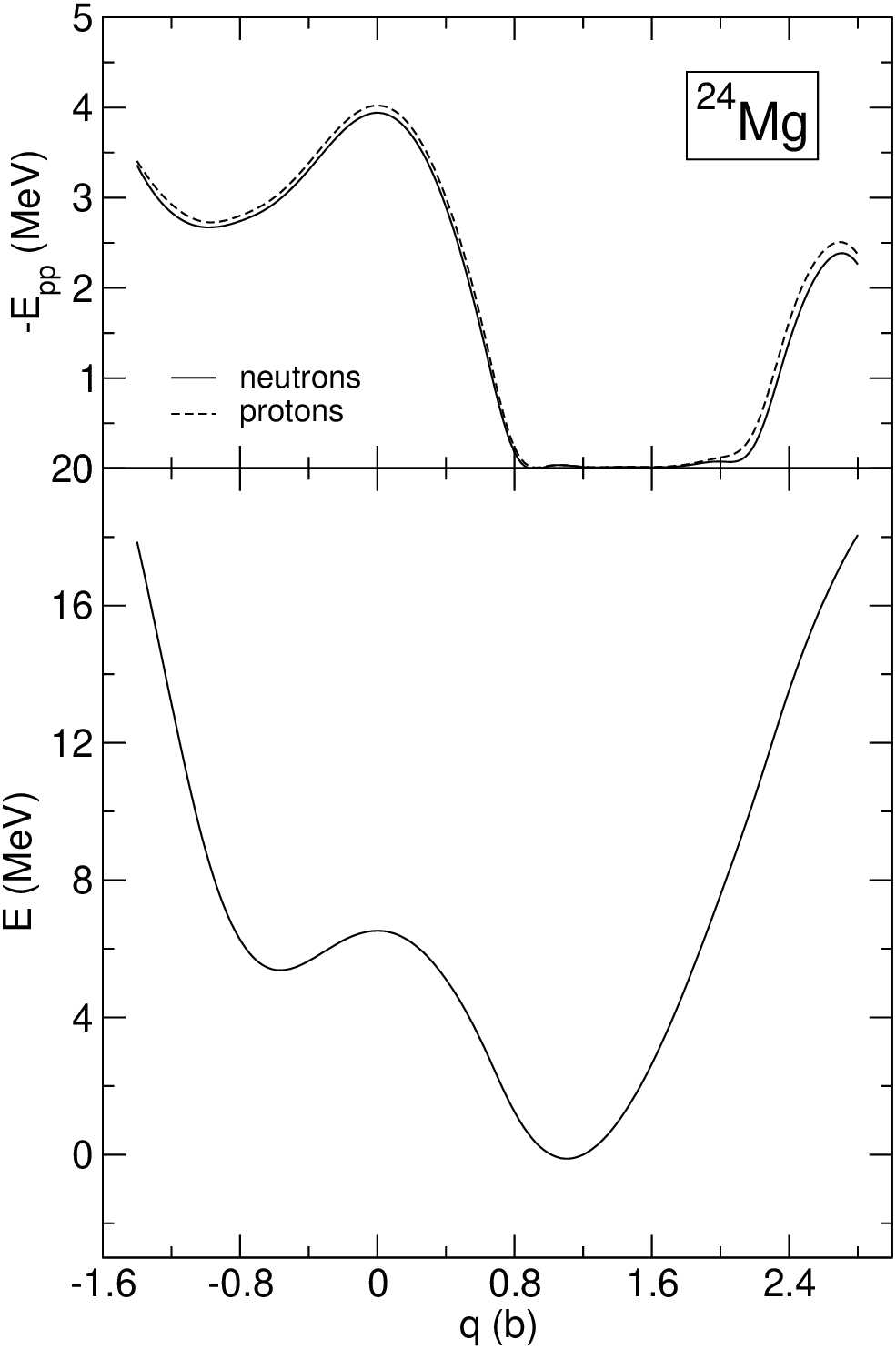}
\caption{The BCS pairing energy for protons and neutrons (upper panel),
and the mean-field plus BCS binding energy curve (lower panel)
of $^{24}$Mg, as functions of the mass quadrupole moment.}
\label{figB}
\end{figure}

In Fig.~\ref{figB} we display the pairing energy (upper panel), and the
total RMF binding energy curve (lower panel) of $^{24}$Mg, as functions of
the mass quadrupole moment. The Lipkin-Nogami procedure has not been
implemented at this stage, and consequently pairing correlations vanish
in a broad region of deformations around the deformed first minimum of the
potential energy surface. Since the moment of inertia
of a rotational band is reduced in the presence of pairing
correlations, dynamical pairing effects could be important in
the description of the ground-state band of $^{24}$Mg.

\begin{figure}
\includegraphics[scale=0.6]{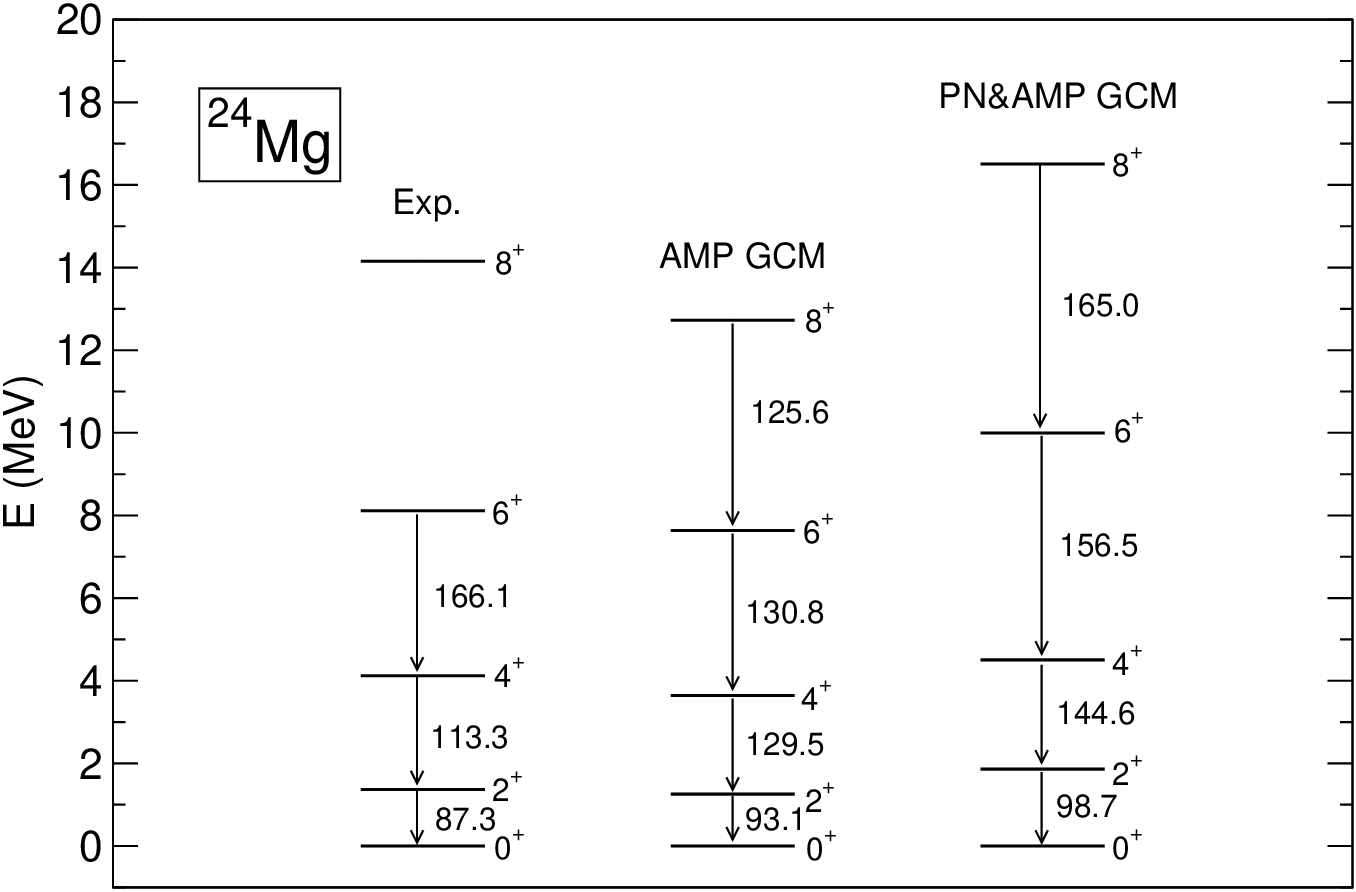}
\vspace{1cm}
\caption{The ground-state rotational band in $^{24}$Mg. The B(E2) values
are in units of $e^2fm^4$. The GCM spectra calculated with (PN\&AMP),
and without (AMP) particle number projection, are compared with the
experimental ground-state band~\cite{Endt.90}}
\label{figC}
\end{figure}

The GCM excitation energies and the resulting transition
probabilities for the ground-state band, calculated with the PC-F1
effective interaction, are shown in Fig.~\ref{figC}. The results of
the AMP and PN\&AMP configuration mixing calculations are compared
with the data. As expected, the inclusion of dynamical pairing
effects reduces the moment of inertia, but the resulting spectrum is
much too spread out compared to experiment. This is a well known
problem, related to the fact that we project particle number and
angular momentum only after variation, rather than performing the
projections before variation~\cite{VS.71}. It has been shown that in
the latter case rotational bands with larger moments of inertia are
obtained \cite{HHR.82,SG.87}, provided that the model geometry
allows for the alignment of nucleon angular momenta. Since the full
projection before variation is technically and computationally much
more complex, it has been seldom used in realistic calculations. We
note that one possible improvement of the present model would be to
project, for each value of the angular momentum $J$, the cranked
mean-field wave functions which, in addition to the mass quadrupole
moment, are also constrained to have  $\langle
J_x\rangle=J$~\cite{RS.80,BH.84}. This extension has not been
included in the present analysis.

The transition probabilities, as well as the calculated
spectroscopic quadrupole moment $Q_{spec}(2_1^+)=-16.56$ e fm$^2$,
are in very good agreement with the data. The AMP GCM results are
very similar to those obtained by using the Gogny D1S force (see
Figs. 11 and 13 in Ref.~\cite{RER.02b}), whereas the PN\&AMP
spectrum is close to the one calculated with the Skyrme SLy4
interaction (see Fig. 6 in Ref.~\cite{VHB.00}).

\begin{figure}
\includegraphics[scale=0.6]{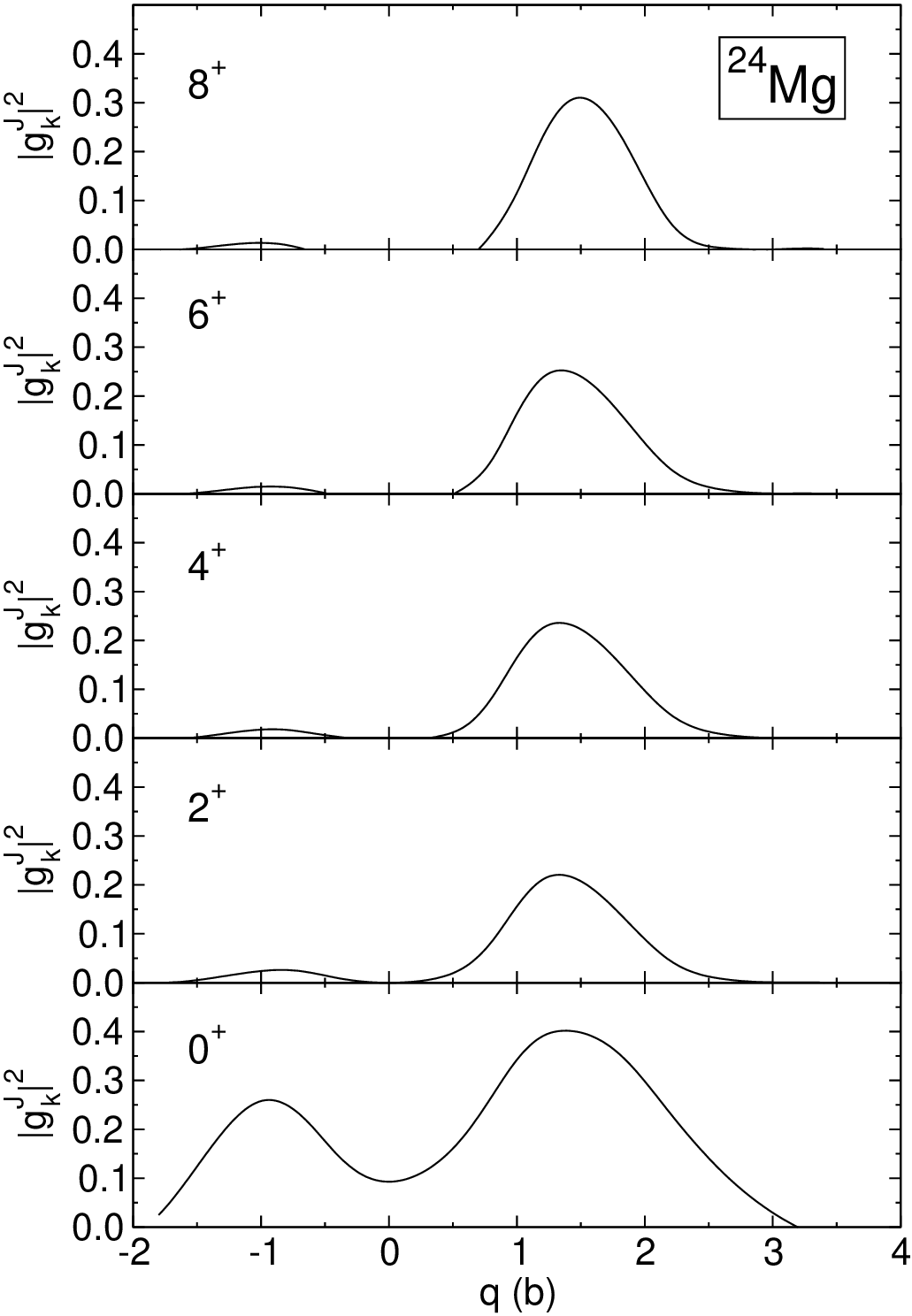}
\caption{The amplitudes of the PN\&AMP GCM collective wave
functions $|g_k^J|^2$ of the ground-state band in $^{24}$Mg.}
\label{figD}
\end{figure}

The amplitudes of the PN\&AMP GCM collective wave functions $|g_k^J|^2$
of the ground-state band, are plotted in Fig.~\ref{figD} as
functions of the mass quadrupole moment. It is interesting to note
that only the $0_1^+$ state contains significant admixtures of oblate
deformed shapes, whereas the amplitudes of states with $J\ge 2$ are
concentrated in the prolate well. The same result has also been
obtained with the Skyrme SLy4 interaction
(see Fig. 4 of Ref.~\cite{VHB.00}).

\subsection{\label{subIIIb}$^{32}$S}

In recent years a number of theoretical and experimental studies of
the structure of $^{32}$S have been reported. This nucleus is
particularly interesting because the excitation energies of the
low-lying $0_2^+$, $2_1^+$ and $4_1^+$ states correspond to those of
a typical spherical vibrator, whereas on the other hand, the
quadrupole moment of the first $2^+$ state is large and negative,
indicating large dynamical prolate deformation \cite{Rag.89}.
$^{32}$S is among the best studied nuclei in the $sd$ shell, and
both the energies and lifetimes of many states up to $\approx 10$
MeV excitation energy are known \cite{Kan.98,BHM.02}.

Several modern theoretical approaches have been used in recent
studies of normally deformed (ND), and superdeformed (SD)
configurations in $^{32}$S: the shell model with the universal
sd-shell Hamiltonian \cite{Bre.97}, the semimicroscopic algebraic
cluster model \cite{CLV.98}, and various extensions of the
self-consistent mean-field framework. They include the cranked
Hartree-Fock \cite{MDD.00,YM.00,TNI.01}, and Hartree-Fock-Bogoliubov
method \cite{RER.00} for the description of the SD configuration,
and the generator coordinate method with the Skyrme
SLy6~\cite{BFH.03}, and Gogny D1S effective interactions, in the
analysis of both ND and SD configurations~\cite{RER.00,BFH.03}.

\begin{figure}
\includegraphics[scale=0.6]{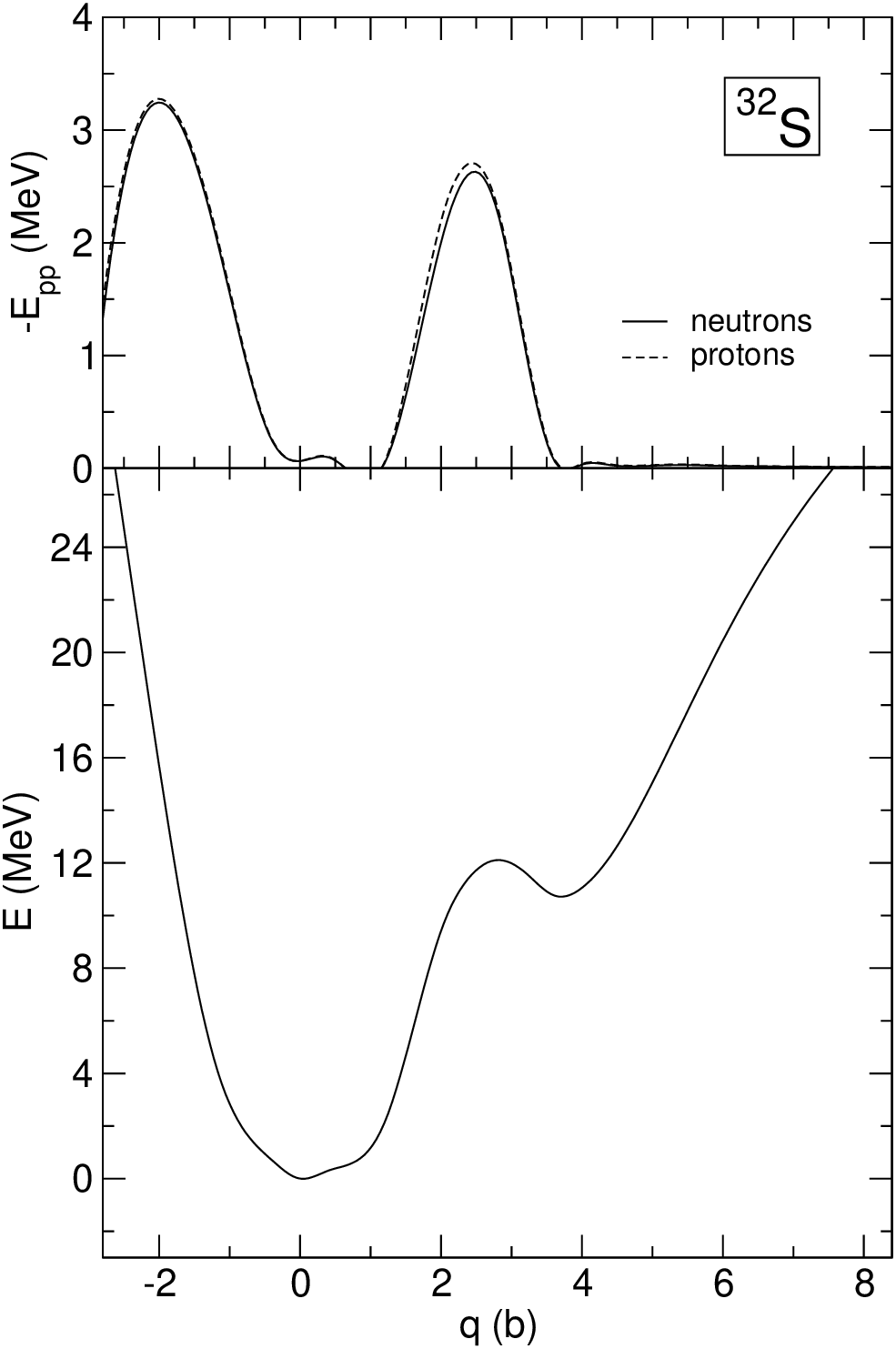}
\caption{The BCS pairing energy for protons and neutrons (upper panel),
and the mean-field plus BCS binding energy curve (lower panel)
of $^{32}$S, as functions of the mass quadrupole moment.}
\label{figE}
\end{figure}

In Fig.~\ref{figE} the pairing energy (upper panel), and the
total RMF binding energy curve (lower panel) of $^{32}$S, are
plotted as functions of the mass quadrupole moment. The effective
interaction is PC-F1, and pairing correlations are treated in the
BCS approximation, without particle number projection.
Consequently, we notice the collapse of the pairing energy both
in the ND and the SD minima. This means that restoring the
particle number symmetry could have an important effect in
the description of both ground-state and SD bands.

\begin{figure}
\includegraphics[scale=0.7]{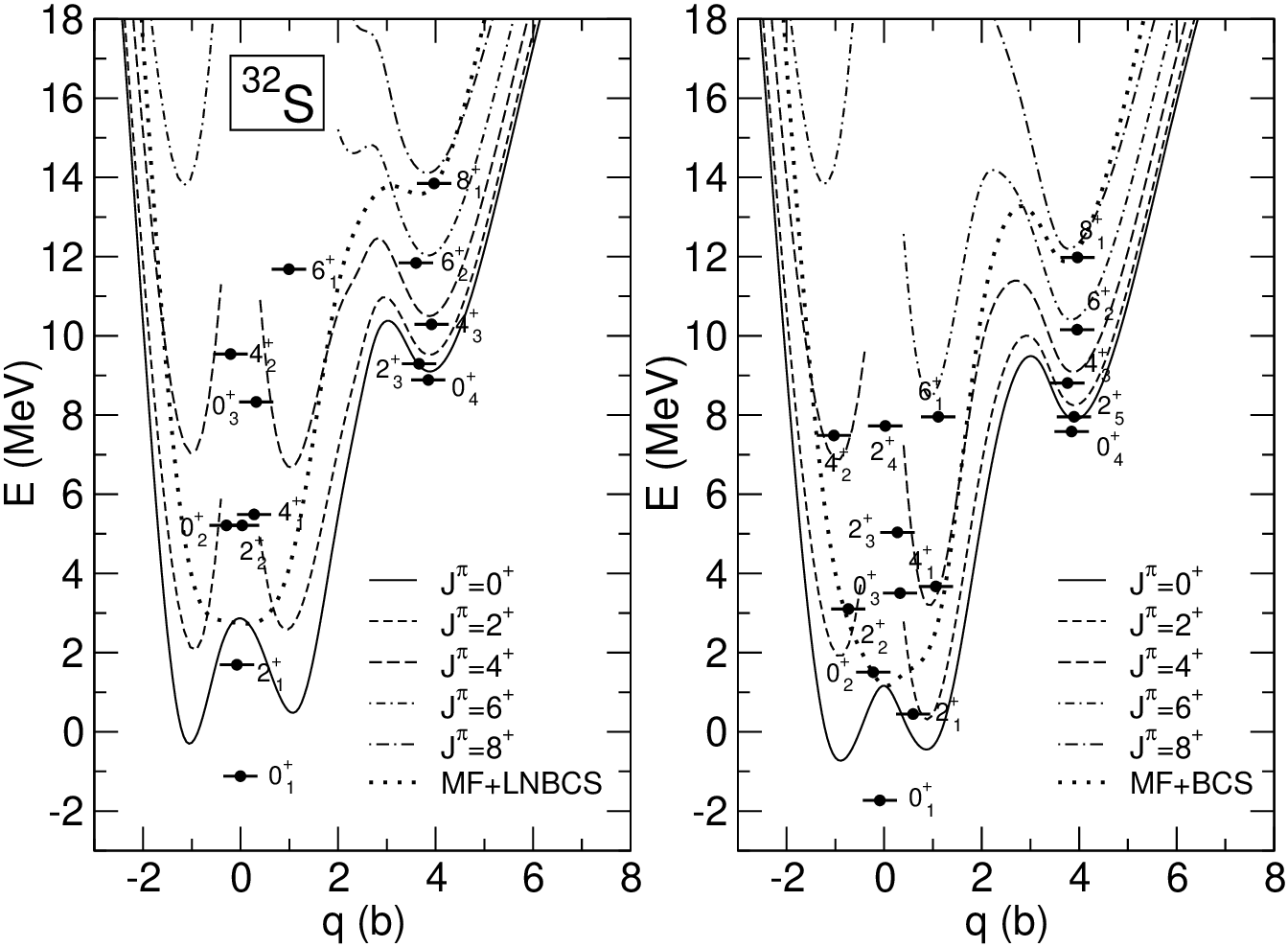}
\caption{The energies and the average quadrupole moments of the GCM states in
$^{32}$S, plotted together with the corresponding angular momentum
projected energy curves. GCM calculations
with (left panel), and without (right panel) particle number projection are
compared. The mean-field binding energy curves are also included in the figure
(dotted curves). In both panels zero energy corresponds to the
minimum of the $J=0$ projected energy curve.}
\label{figF}
\end{figure}

The results of the PN\&AMP and AMP GCM calculations for $^{32}$S are
shown in the left and right panels of Fig.~\ref{figF}, respectively.
In the left (right) panel we also include the mean-field MF+LNBCS (MF+BCS)
binding energy curve (dotted curves). In both cases the MF energy curve
displays a spherical minimum, and an additional shallow minimum at
large deformation $q\approx 4$b with excitation energy $E_x \approx 11$ MeV.
The occurrence of almost degenerate oblate and prolate minima
in the $J=0^+$ projected energy curve, symmetrical with
respect to the spherical configuration, is a feature common to all nuclei
for which the mean-field calculation predicts a spherical ground
state (see, for instance, Refs.~\cite{BFH.03,ER.04}).

The superdeformed minimum is more pronounced in the calculation
without particle number projection, both for the MF+BCS curve and
for the angular momentum projected energy curves. The GCM superdeformed
band is calculated at somewhat lower excitation energies in the AMP case,
and this is because in the PN\&AMP calculation pairing correlations
do not vanish in the SD minimum. The energies of the GCM states
are plotted as functions of the average quadrupole moment
\begin{equation}
q=\langle q_k \rangle=\sum_j{g_k^2(q_j)q_j}\;. \label{quadav}
\end{equation}
The overall structure of energy levels, and in particular the
two-phonon triplet $0_2^+$, $2_2^+$, $4_1^+$, is much better described when
particle number projection is included (see Tables \ref{TabA} and
\ref{TabB} for a comparison with experimental levels).
Similar results for the
spectra of $^{32}$S have also been obtained in the GCM analyses of
Refs.~\cite{BFH.03,RER.00}. The calculation with
the zero-range SLy6 effective interaction, including particle number
projection, reproduces the structure of the two-phonon triplet \cite{BFH.03}.
On the other hand, the results for the ND states obtained with
the finite-range Gogny force, but with only angular momentum
projection \cite{RER.00}, are similar to those shown in the right panel
of Fig.~\ref{figF}, i.e. an additional low-lying $0^+$ state is
predicted by the calculation.

In Tab.~\ref{TabA} we list the PN\&AMP GCM excitation energies, the
spectroscopic quadrupole moments, and the E2 (J $\to$ J -2) transition
probabilities for the  $2_1^+$ state, and for the two-phonon triplet
$0_2^+$, $2_2^+$, $4_1^+$, in comparison with the
results of Ref.~\cite{BFH.03}, and the available data. The results obtained
with both effective interactions, the nonrelativistic SLy6 and the relativistic
PC-F1, are in qualitative agreement with the data. The excitation energies of
the two-phonon triplet are calculated at approximately twice the
energy of the $2_1^+$ state. We note, however, that the
predicted quadrupole moments for the $2_1^+$ state are small and positive,
whereas the experimental value points to a much larger and prolate
deformation of this state.

A small prolate deformation for the ground-state band is only obtained in
calculations which do not include particle number projection (right panel
of Fig.~\ref{figF}). In Tab.~\ref{TabB} we compare the AMP GCM excitation
energies and B(E2, J $\to$ J -2) values obtained with PC-F1,
with the corresponding quantities calculated
with the Gogny D1S interaction \cite{RER.00}. The results are very
similar, with the largest differences in the B(E2)
values for the transitions $2_1^+\to 0_1^+$ and $2_2^+ \to 0_3^+$. This
can be attributed to the smaller values of
the quadrupole moments of the $2^+$ states predicted by the
PC-F1 interaction. For instance, the spectroscopic quadrupole moment of the
$2_1^+$ state calculated with the Gogny force $-13.29$ e fm$^2$
is close to the experimental value: $-14.9$ e fm$^2$~\cite{Endt.90},
whereas the one predicted by the PC-F1 interaction: $-3.5$ e fm$^2$, is
much too small.

\begin{figure}
\includegraphics[scale=0.6]{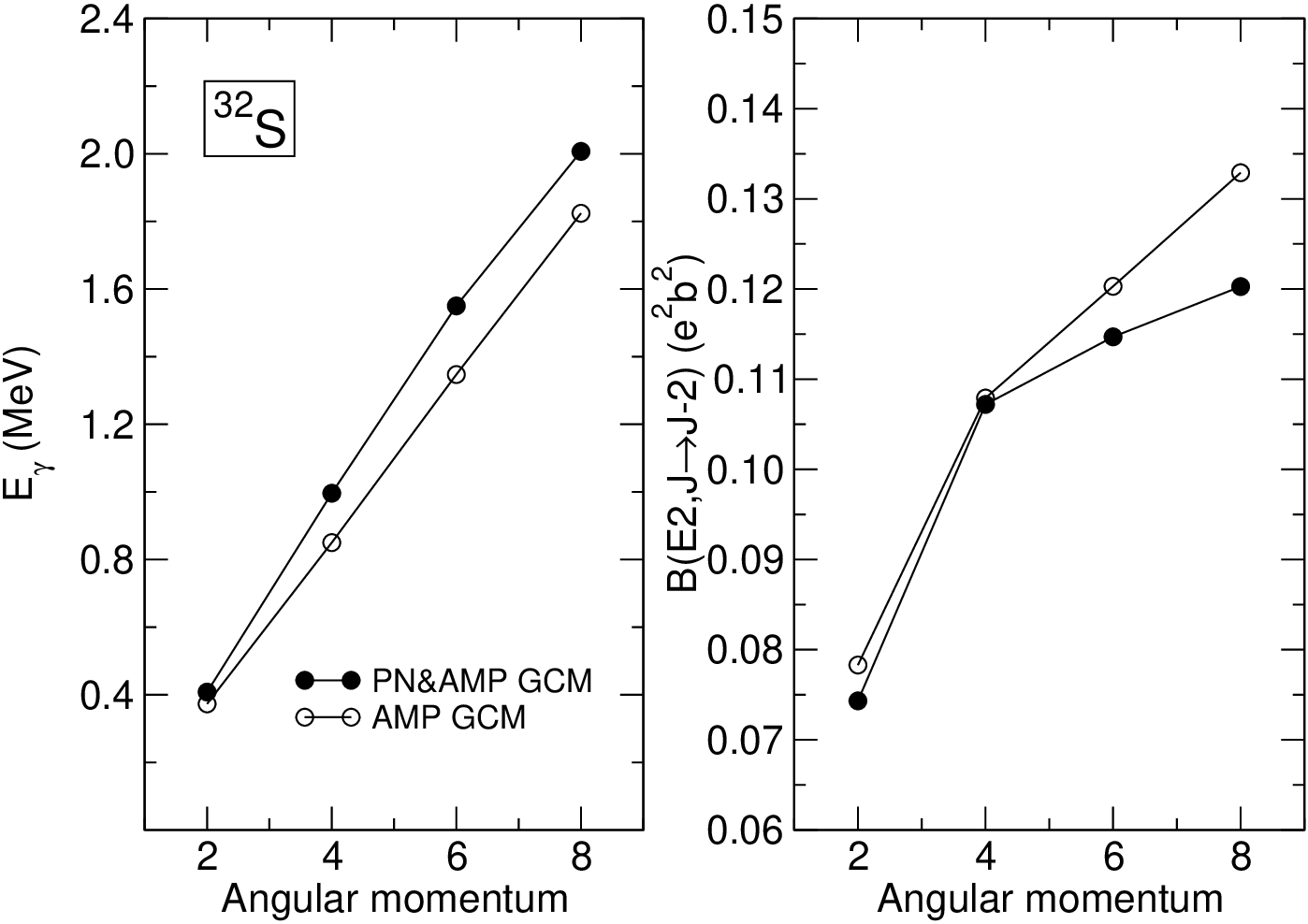}
\vspace{1cm}
\caption{Transition energies
$\Delta E (J)= E(J)-E(J-2)$ (left panel), and the B(E2, J$\to$ J-2)
values (right panel), for the SD configuration in
$^{32}$S, as functions of the angular momentum.}
\label{figG}
\end{figure}

Although the present version of the RMF plus GCM model is not
optimal for the study of moments of inertia of superdeformed rotational bands,
because it does not include cranked wave functions, nevertheless it can be used
to investigate the stability of the SD intrinsic configurations against
quadrupole fluctuations at low angular momentum. In Fig.~\ref{figF}
it is shown that both the PN\&AMP and AMP GCM calculations predict an
SD band at large deformation $q\approx 4b$. In the left panel of
Fig.~\ref{figG} we plot the energy differences
$\Delta E (J)= E(J)-E(J-2)$, as functions of the angular momentum of
the SD band. Since without particle number projection pairing
correlations vanish in the SD minimum, the AMP GCM
calculation overestimates the moment of inertia of the SD band, and the
resulting spectrum is more compressed. The moment of inertia is
reduced with the inclusion of dynamical pairing effects.
In the right panel of Fig.~\ref{figG} we include the corresponding
E2 transition probabilities. Both the relative excitation energies
and the B(E2) values for the SD band are consistent with the
results reported in Ref.~\cite{BFH.03} (Tab. VIII), and
Ref.~\cite{RER.00} (Fig. 3). We note, however, differences in
the predicted  position of the SD band-head. In our
PN\&AMP GCM calculation the SD band-head is found at $8.9$ MeV,
almost $3$ MeV lower than the value calculated with the SLy6 interaction in
Ref.~\cite{BFH.03}. The AMP GCM calculation predicts the head of the
SD band at $7.6$ MeV, comparable to the value obtained with the Gogny D1S
interaction in Ref.~\cite{RER.00} ($8.9$ MeV). A possible reason for the
large difference between the PC-F1 and Gogny D1S interactions on one hand,
and the SLy6 Skyrme force on the other, can be identified already at
the mean-field level. While the energy curves obtained with the
PC-F1 and Gogny interactions (see also Fig. 2 in Ref.~\cite{RER.00})
exhibit a second, superdeformed shallow minimum at $q\approx 4b$, only a
shoulder in the potential energy curve at superdeformation is predicted
by the SLy6 interaction  (see Fig. 13 in Ref.~\cite{BFH.03}).

\subsection{\label{subIIIc}$^{36}$Ar}
In the last example we present the GCM analysis of the low angular
momentum structure of $^{36}$Ar, one of the lightest nuclei in which
a superdeformed structure has been studied experimentally. The
excitation energies and $E2$ transition probabilities for the
prolate SD band have been recently measured up to the angular
momentum $I=16\hbar$~\cite{Sve.00}. The properties of the oblate
ground-state band were determined over a decade ago~\cite{Endt.90}.
A number of theoretical analyses of the structure of $^{36}$Ar
include the cranked Nilsson-Strutinsky and the shell
model~\cite{Sve.00,Sve.01}, the projected shell model~\cite{LS.01},
the self-consistent cranked Hartree-Fock-Bogoliubov
model~\cite{RER.04}, and the generator coordinate method with Skyrme
SLy6~\cite{BFH.03} and Gogny D1S~\cite{RER.04} interactions.

\begin{figure}
\includegraphics[scale=0.7]{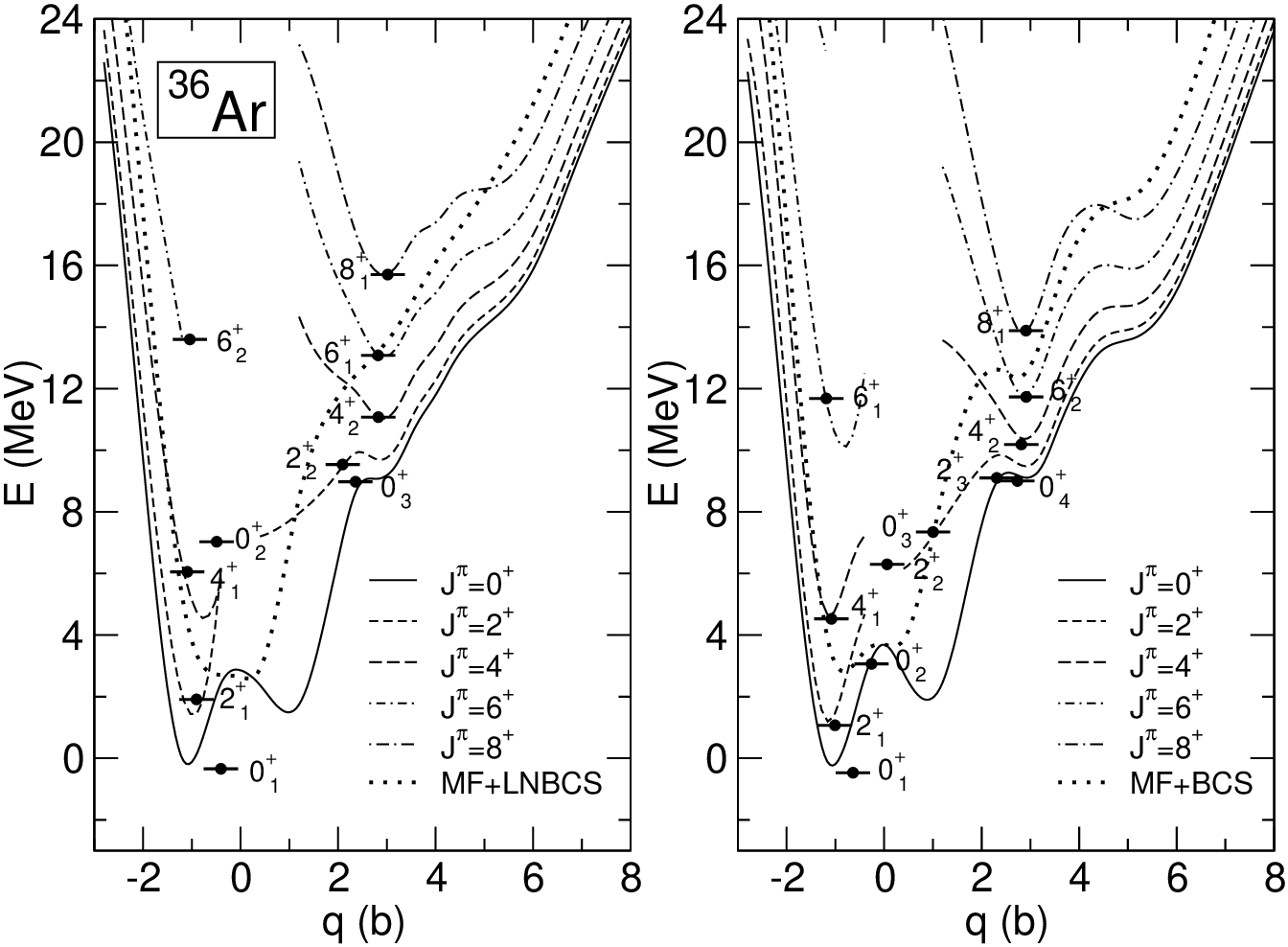}
\caption{The energies and the average quadrupole moments of the GCM states in
$^{36}$Ar, plotted together with the corresponding angular momentum
projected energy curves. GCM calculations
with (left panel), and without (right panel) particle number projection are
compared. The mean-field binding energy curves are also included in the figure
(dotted curves). In both panels zero energy corresponds to the
minimum of the $J=0$ projected energy curve.}
\label{figH}
\end{figure}

The results of the PN\&AMP and the AMP GCM calculations are shown in the
left and right panels of Fig.~\ref{figH}, respectively.
In both cases the mean-field binding energy curve is also included
(dotted curves). The BCS mean-field energy curve displays an
oblate minimum, rather flat in the region of deformation:
$-1$ b $\le$ q $\le 0.5$ b. In addition, a shallow minimum is found
at larger deformation $q \approx 2.8$ b, at $E_x \approx 9$ MeV above
the ground-state minimum, and a shoulder is predicted at
still larger deformation $q\approx 5$ b. A similar mean-field potential
energy surface is obtained with the Gogny D1S interaction
(see Fig. 1 in Ref.~\cite{RER.04}), but the SD minimum is
calculated $\approx 1$ MeV lower than with PC-F1. We note that
in the BCS approximation
pairing correlations vanish both in the ND and SD minimum,
and this means that the restoration of particle number can
produce sizable effects in the ground-state and SD bands.
The mean-field energy curve which includes the approximate
particle number projection (LNBCS), displays only a weak
shoulder, rather than a  minimum at deformation $q\approx 3$ b.
This curve is in qualitative agreement with the LNBCS binding
energy curve calculated with the Skyrme
SLy6 effective interaction (see Fig. 10 in Ref.~\cite{BFH.03}),
although in the latter case the shoulder occurs at smaller deformation
$q\approx 1.8$ b, and much smaller excitation energy $E_x\approx 5$ MeV.
Both in the PN\&AMP and the AMP calculations, the
$J=0$ angular momentum projected energy curves display two well
developed low-lying minima on the oblate and prolate side.
The oblate minimum corresponds to the ground state. An additional SD
minimum occurs on the AMP $J=0$ energy curve, whereas the corresponding
PN\&AMP curve displays only a plateau. The angular momentum projected energy
curves with $J \ge 2$ exhibit well developed SD minima both in the
AMP and PN\&AMP calculations. Finally, the
energies of the resulting GCM states for angular momenta
$J \leq 8$ are included as functions of the average
quadrupole moment, defined in Eq.~(\ref{quadav}). The low-spin GCM
spectra contain an oblate ND ground state band, and a prolate SD band.

In Tab.~\ref{TabC} we compare the PN\&AMP GCM excitation energies, the
spectroscopic quadrupole moments, and the B(E2, J $\to$ J-2) values
for the ground-state band, with the results obtained
with the Skyrme SLy6 interaction in Ref.~\cite{BFH.03}, and with
available data. The theoretical results are in qualitative agreement.
Although the energy spectrum is described somewhat better by the PC-F1
interaction, both calculations clearly underestimate the moment of
inertia of the ND ground-state band. In Tab.~\ref{TabD} the
AMP GCM excitation energies and the B(E2, J $\to$ J-2) values
for the ND band, are shown in comparison with those calculated with the
Gogny D1S interaction in Ref.~\cite{RER.04}. The results obtained with
the PC-F1 and Gogny effective interactions are very similar, and a
considerable increase of the moment of inertia as compared to the
PN\&AMP GCM spectra, is caused by the collapse of pairing correlations
in the ND minimum.

\begin{figure}
\includegraphics[scale=0.6]{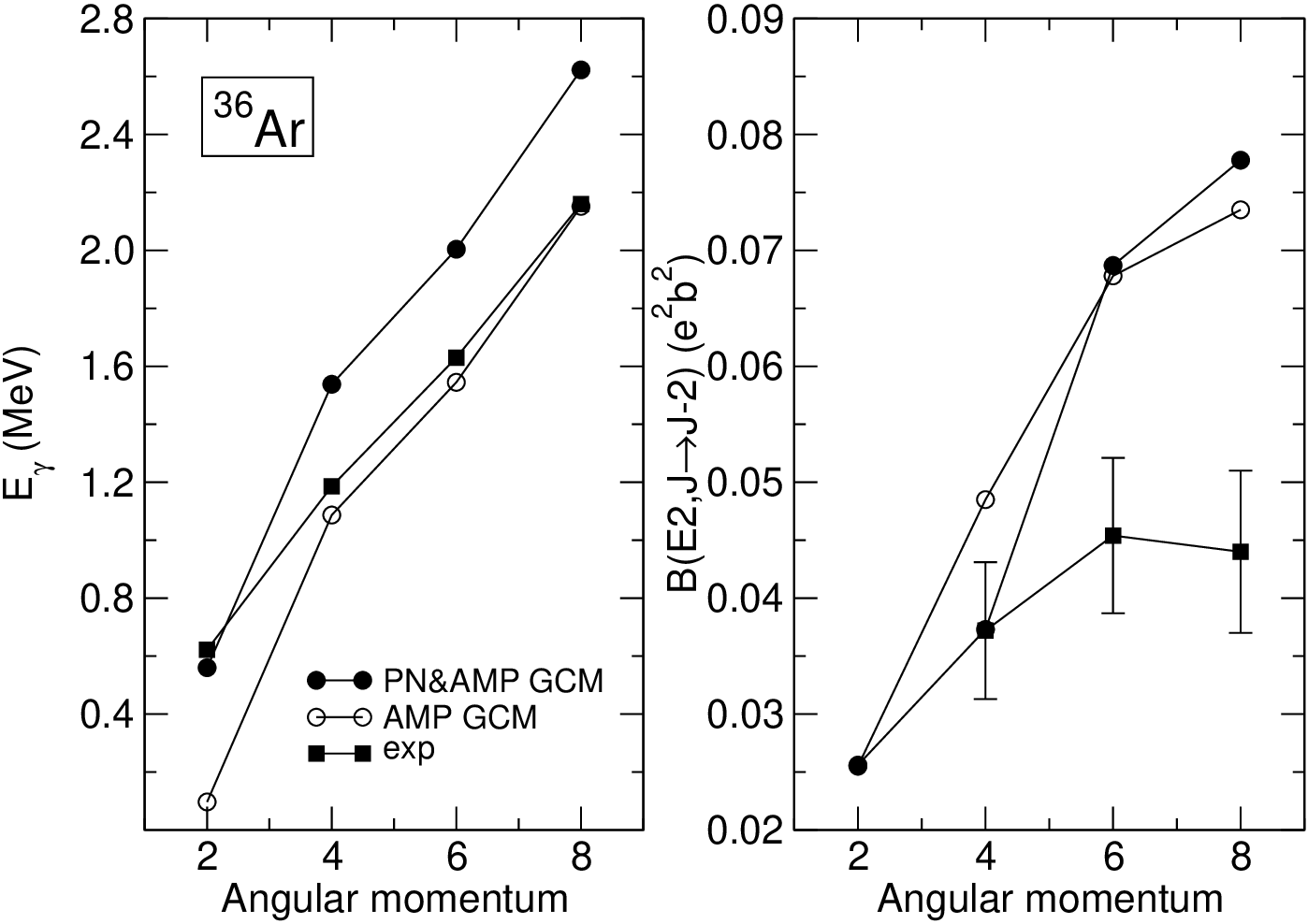}
\caption{The energy differences
$\Delta E (J)= E(J)-E(J-2)$ (left panel), and the B(E2, J$\to$ J-2)
values (right panel), for the SD band in
$^{36}$Ar, as functions of the angular momentum. Results of GCM
calculations with and without particle number projection are
compared with the available data \protect\cite{Sve.00,Sve.01}.}
\label{figI}
\end{figure}

In the left panel of Fig.~\ref{figI} we plot the energy differences
$\Delta E (J)= E(J)-E(J-2)$, and in the right panel the $BE2$ values,
as functions of the angular momentum of the SD band, in comparison with data.
Cranked Hartree-Fock-Bogoliubov calculations performed with the Gogny
interaction \cite{RER.04}, have shown that rather strong triaxiality
effects appear already at zero-spin and, as a result, predicted
a steady decrease of deformation with increasing angular
momentum. Being restricted to axially symmetric shapes, the model that
we have developed in this work  cannot take these effects into account.
This is clearly reflected in the pronounced discrepancy between the calculated
and experimental B(E2) values. Similar results have also been obtained
with the SLy6 (see Tab. V in Ref.~\cite{BFH.03}), and Gogny effective
interactions (see Fig. 3 in Ref.~\cite{RER.04}). Just like in the case
of $^{32}$S that we have considered in the previous section,
the transition energies and the B(E2) values do not crucially
depend on the effective interaction in the particle-hole channel.
On the other hand, the predicted positions of the SD band-head
differ significantly: for the Gogny D1S interaction the
band-head is at $7.5$ MeV, whereas the SLy6 interaction predicts a
lower value of $5.9$ MeV. The positions of SD band-head
obtained in the PN\&AMP and AMP GCM calculations with the PC-F1
interaction are $9.2$ MeV and $9.4$ MeV, respectively.

\begin{figure}
\includegraphics[scale=0.6]{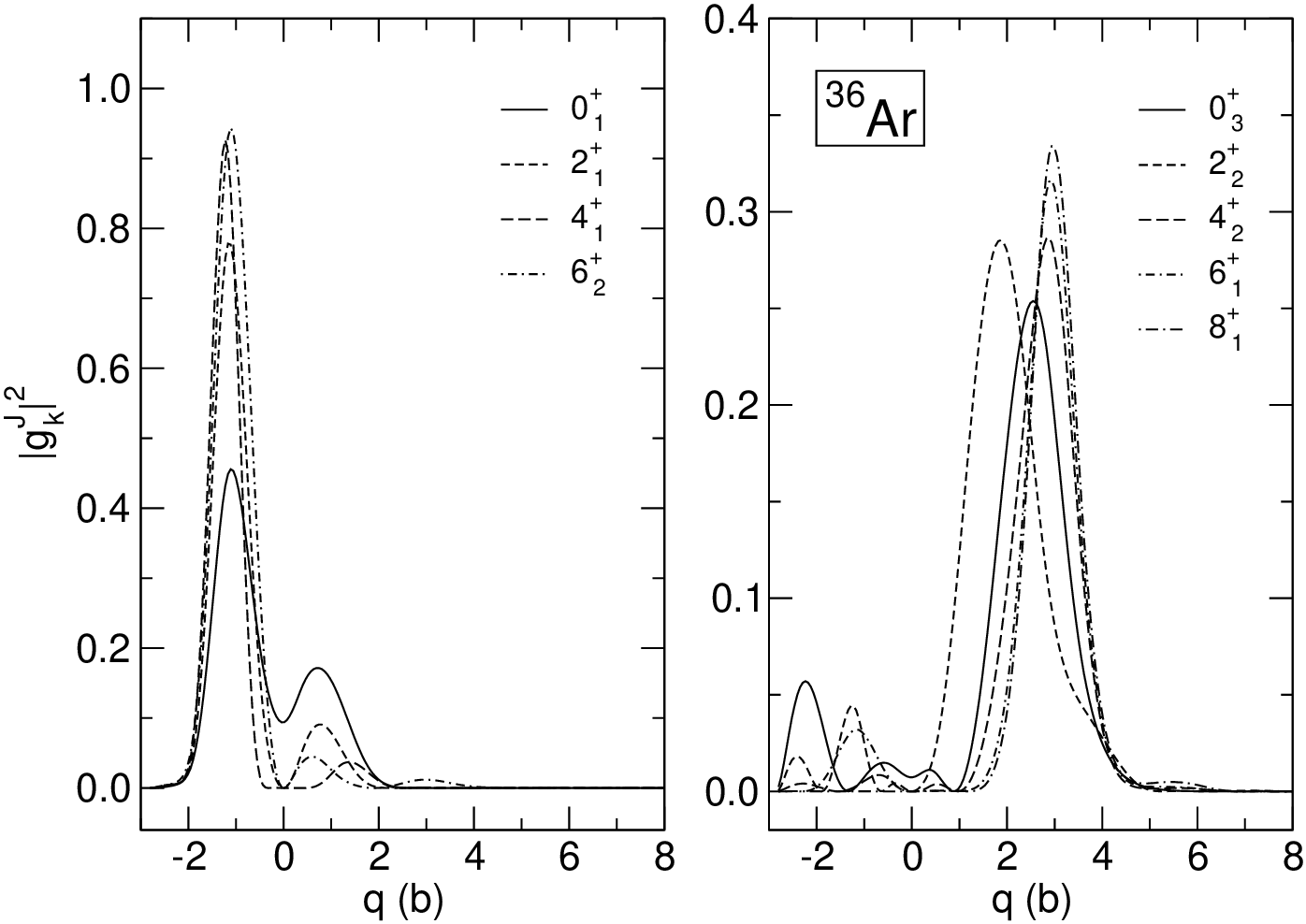}
\vspace{1cm}
\caption{The amplitudes of the PN\&AMP GCM collective wave
functions $|g_k^J|^2$ for the ND ground-state band
($0_1^+$, $2_1^+$, $4_1^+$ and $6_2^+$), and the SD band
($0_3^+$, $2_2^+$, $4_2^+$ and $6_1^+$ and $8_1^+$) in $^{36}$Ar,
as functions of the mass quadrupole moment.}
\label{figJ}
\end{figure}

Finally, the amplitudes of the collective PN\&AMP GCM wave functions
$|g_k^J|^2$ for the
ND ($0_1^+$, $2_1^+$, $4_1^+$ and $6_2^+$), and the SD
($0_3^+$, $2_2^+$, $4_2^+$ and $6_1^+$ and $8_1^+$) rotational bands in
$^{36}$Ar are plotted in Fig.~\ref{figJ}. We notice that, except for
the ground state $0_1^+$, the amplitudes of the states of the ND band are
principally concentrated in the oblate minimum, whereas in the panel on
the right the amplitudes of the wave functions are strongly peaked in
the prolate superdeformed minimum.
\section{\label{secIV}Summary and outlook}
In Ref.~\cite{NVR.06} and in this work we have extended the very successful
relativistic mean-field theory \cite{BHR.03,VALR.05} to
explicitly include correlations related to the restoration of broken
symmetries and to fluctuations of collective coordinates. We have
developed a model that uses the generator coordinate method to
perform configuration mixing of angular-momentum and
particle-number projected relativistic wave functions.
The geometry is restricted to axially symmetric shapes, and
the intrinsic wave functions are generated from the solutions of
the relativistic mean-field + Lipkin-Nogami
BCS equations, with a constraint on the mass quadrupole moment.
The single-nucleon Dirac eigenvalue equation is solved by
expanding the large and small components of the nucleon spinor
in terms of eigenfunctions of an axially symmetric harmonic
oscillator potential.
The current implementation of the model employs
a relativistic point-coupling (contact)
nucleon-nucleon effective interaction in the particle-hole
channel, and a density-independent $\delta$-interaction in the
particle-particle channel.

We have performed several illustrative calculations
which test our implementation of the GCM with simultaneous
particle-number and angular-momentum projection. The results
have been compared with those obtained employing the
relativistic GCM model developed in the first part of this work
\cite{NVR.06}, which does not include particle-number projection,
but only conserves the number of particles on the average, and
with results that were obtained using the corresponding
non-relativistic models based on Skyrme
and Gogny effective interactions. In this work the low-lying
spectra of $^{24}$Mg, $^{32}$S and $^{36}$Ar
have been analyzed. The results of GCM configuration
mixing calculations both without and with particle-number
projection, have been compared with those obtained with the
angular-momentum projected GCM based on the Gogny D1S
effective interaction, and with the particle-number and angular-momentum
projected GCM based on the Skyrme SLy4 and SLy6 effective interactions.
The principal features of the spectra calculated without
(Gogny interaction) and with particle-number projection (Skyrme forces),
are very well reproduced in our GCM calculation with the PC-F1
relativistic effective interaction. These include the deformations and
moments of inertia of yrast bands, and the occurrence and structure
of superdeformed bands. Some differences between the predictions of
the three models, as for instance the position of the head
$0^+$ of the superdeformed band, can be attributed to the gaps
in the mean-field single-nucleon spectra calculated with the
different effective interactions. We have also shown that dynamical
pairing effects play an important role in the description of the
low-energy spectra. In particular, we find a pronounced effect
of particle-number projection on the moments of inertia of the
ground-state and superdeformed rotational bands. Another example is
the two-phonon triplet $0_2^+$, $2_2^+$ and $4_1^+$ in
$^{32}$S, which can be reproduced by GCM calculations only when
dynamical pairing effects are taken into account.

There are, of course, many possible improvements and extensions of
the present implementation of the relativistic GCM model. Perhaps
the most obvious is the extension to shapes that are not constrained
by axial symmetry. The inclusion of triaxial deformations is in
principle straightforward but, because it requires an enormous
increase of computational capabilities, not feasible at present. The
second major problem is that our GCM configuration mixing
calculations correspond to a projection after variation. A more
general variation after projection is far too complicated to be used
in realistic calculations at the present stage. A possible
improvement, however, is to generate the GCM basis functions, for
each value of the angular momentum, by performing cranking RMF+LNBCS
calculations with the additional constraint $\langle
J_x\rangle=\sqrt{J(J+1)}$. This would automatically increase the
moments of inertia of rotational bands, and therefore produce
spectra in better agreement with experiment. The extension to odd-A
nuclei necessitates the breaking of time-reversal invariance in the
wave functions and, therefore, the explicit inclusion of currents in
the energy-density functional, and the corresponding time-odd fields
in the single-nucleon Dirac equation. Finally, let us emphasize
again one the conclusions of Ref.~\cite{NVR.06}, namely that those
correlations which are explicitly treated in the GCM configuration
mixing, should not be contained in the effective interaction in an
implicit way, i.e. by adjusting the parameters of the interaction to
data which already include these correlations. Therefore, before the
present version of the relativistic self-consistent GCM is applied
in realistic calculations, we need to adjust a new global effective
interaction which will not contain symmetry breaking corrections and
quadrupole fluctuation correlations.
\appendix*
\section{\label{appI}Evaluation of the parameter $\lambda_2$ in the
Lipkin-Nogami method}
The variation of the auxiliary functional $\mathcal{K}$
Eq.~(\ref{LNfunctional}) can be expressed as
\begin{equation}
\langle \mathcal{K} \hat{N}_2^2\rangle=0\;,
\label{Kvar}
\end{equation}
for the ground-state expectation value, with
$\hat{N}_2$ as the two-quasiparticle part of the particle-number
operator Eq.~(\ref{n2}). Since we do not consider proton-neutron pairing,
the equations for protons and neutrons separate, and the isospin
index $q$ can be omitted in the following derivation.
If one introduces the shifted many-body state
\begin{equation}
\ket{\xi}=e^{i\xi\hat{N}_2}\ket{0},\qquad \ket{0}=\ket{\xi}\Big{\vert}_{\xi=0}\;,
\label{shift}
\end{equation}
the following expression for $\lambda_2$ is obtained from the
variational condition Eq.~(\ref{Kvar}) \cite{RNB.96}
\begin{equation}
\lambda_2 = \frac{\partial_\xi^2 \bra{0}\hat{H}\ket{\xi}\Big{\vert}_{\xi=0}}
          {\partial_\xi^2 \bra{0}\hat{N}^2\ket{\xi}\Big{\vert}_{\xi=0}}\;.
\label{lambda2}
\end{equation}
The denominator of this equation is evaluated using Wick's theorem
\begin{equation}
\partial_\xi^2 \bra{0}\hat{N}^2\ket{\xi}\Big{\vert}_{\xi=0}=32\left[
    (\overline{u^2v^2})^2-\overline{u^4v^4}\right]\;.
\label{denom}
\end{equation}
The principal advantage of this particular formulation of the
Lipkin-Nogami scheme is that one can relate it to the theoretical
framework of energy density functionals~\cite{RNB.96}
\begin{equation}
\bra{0}\hat{H}\ket{\xi} \Longleftrightarrow \mathcal{E}^{(\xi)}=
   \mathcal{E}[\hat{\rho}^{(\xi)},\hat{\kappa}^{(\xi)},\hat{\kappa}^{*(\xi)}]\;,
\end{equation}
with the following definition of the shifted densities
\begin{equation}
\rho_{kl}^{(\xi)}=\bra{0}a_l^\dagger a_k\ket{\xi}, \qquad
\kappa_{kl}^{(\xi)}=\bra{0}a_{\bar{l}} a_k\ket{\xi}\;.
\end{equation}
The second derivative of the energy functional reads
\begin{eqnarray}
\partial_{\xi}^2\mathcal{E}^{(\xi)}&=&{\rm Tr}\left[
   \frac{\delta\mathcal{E}}{\delta\hat{\rho}}\partial_\xi^2\hat{\rho} +
   \frac{\delta\mathcal{E}}{\delta\hat{\kappa}}\partial_\xi^2\hat{\kappa} +
   \frac{\delta\mathcal{E}}{\delta\hat{\kappa}^*}\partial_\xi^2\hat{\kappa}^*\right]
 \nonumber \\
&+&2{\rm Tr}~{\rm Tr} \left[
  \frac{1}{2}\frac{\delta^2\mathcal{E}}{\delta\hat{\rho}_1 \delta\hat{\rho}_2}
     \partial_\xi\hat{\rho}_1 \partial_\xi\hat{\rho}_2 +
   \frac{\delta^2\mathcal{E}}{\delta\hat{\kappa} \delta\hat{\kappa}^*}
      \partial_\xi\hat{\kappa} \partial_\xi\hat{\kappa}^* +
   \frac{\delta^2\mathcal{E}}{\delta\hat{\rho} \delta\hat{\kappa}}
      \partial_\xi\hat{\rho} \partial_\xi\hat{\kappa} +
   \frac{\delta^2\mathcal{E}}{\delta\hat{\rho} \delta\hat{\kappa}^*}
      \partial_\xi\hat{\rho} \partial_\xi\hat{\kappa}^*\right]\;
\label{secder}
\end{eqnarray}
where the trace implies integration and summation over all
coordinates. The term with a single trace vanishes in the BCS ground
state~\cite{RNB.96} and, since we use volume pairing in the energy
functional (see Eq.~\ref{epair}), the mixed derivative terms which
contain $\hat{\rho}$ and $\hat{\kappa}$, or $\hat{\rho}$ and
$\hat{\kappa}^*$, also vanish. With the definition of the response
densities $\tilde{\rho}$, $\tilde{\kappa}$ and $\tilde{\kappa}^*$
\begin{equation}
\tilde{\rho}=-i\partial_\xi\hat{\rho}^{(\xi)}\Big{\vert}_{\xi=0}, \quad
\tilde{\kappa}=-i\partial_\xi\hat{\kappa}^{(\xi)}\Big{\vert}_{\xi=0}
 \quad {\textrm{and}} \quad
\tilde{\kappa}^*=-i\partial_\xi\hat{\kappa}^{(\xi)*}\Big{\vert}_{\xi=0}\;,
\label{respden}
\end{equation}
one finally obtains
\begin{equation}
\partial_{\xi}^2\mathcal{E}^{(\xi)}=-{\rm Tr}~{\rm Tr}\left[
  \frac{\delta^2\mathcal{E}}{\delta\hat{\rho}_1 \delta\hat{\rho}_2}
    \tilde{\rho}_1 \tilde{\rho}_2
+2\frac{\delta^2\mathcal{E}}{\delta\hat{\kappa} \delta\hat{\kappa}^*}
      \tilde{\kappa} \tilde{\kappa}^* \right] \;.
\label{secderfin}
\end{equation}
The RMF energy functional considered in this work consists of three terms:\\
i) the kinetic energy
\begin{equation}
\mathcal{E}_{kin}=\sum_k{\int d\bm{r}~v_k^2~{\bar{\psi}_k (\bm{r})
\left( -i\bm{\gamma}
 \bm{\nabla} + m\right)\psi_k(\bm{r})}}\;,
\label{ekinetic}
\end{equation}
ii) the field energy
\begin{eqnarray}
\mathcal{E}_{field}&=& \int d{\bm r }~{\left(\frac{\alpha_S}{2}\rho_S^2
    +\frac{\beta_S}{3}\rho_S^3 +
  \frac{\gamma_S}{4}\rho_S^4+\frac{\delta_S}{2}\rho_S\triangle \rho_S
 + \frac{\alpha_V}{2}j_\mu j^\mu + \frac{\gamma_V}{4}(j_\mu j^\mu)^2 +
       \frac{\delta_V}{2}j_\mu\triangle j^\mu \right.} \nonumber \\
 &+& \left .
  \frac{\alpha_{TV}}{2}j^{\mu}_{TV}(j_{TV})_\mu+\frac{\delta_{TV}}{2}
    j^\mu_{TV}\triangle  (j_{TV})_{\mu}
 + \frac{\alpha_{TS}}{2}\rho_{TS}^2+\frac{\delta_{TS}}{2}\rho_{TS}\triangle
      \rho_{TS} \right)\;,
\label{efield}
\end{eqnarray}
and iii) the Coulomb interaction term
\begin{equation}
\mathcal{E}_{C}=\frac{e^2}{2}\int\int d{\bm r }d{\bm r^\prime }
       \frac{\rho_p(\bm r)\rho_p(\bm r^\prime)}{| \bm{r}-\bm{r^\prime}|^2}\;.
\label{ecoul}
\end{equation}
The response densities which appear in Eq.~(\ref{secderfin}) are given by
\begin{eqnarray}
\rho_S^\tau(\bm{r})&=&\sum_k u_k^2v_k^2\bar{\psi}_k(\bm{r})\psi_k(\bm{r}) \\
j_\mu^\tau(\bm{r})&=&\sum_k u_k^2v_k^2
            \bar{\psi}_k(\bm{r})\gamma_\mu\psi_k(\bm{r})\;,
\label{respden1}
\end{eqnarray}
where the summation runs for $\tau=n(p)$ over neutron (proton)
single-particle states. The functional derivative of
$\mathcal{E}_{field}$ reads
\begin{eqnarray}
\partial_{\xi}^2\mathcal{E}^{(\xi)\tau}_{field} &=& \int d\bm{r} \left[
   (\alpha_S + 2\beta_S \rho_S + 3\gamma_S \rho_S^2 +
   \alpha_{TS})\tilde{\rho_S}^\tau \tilde{\rho_S}^\tau +
   (\delta_S+\delta_{TS})\tilde{\rho_S}^\tau \triangle \tilde{\rho_S}^\tau \right.\nonumber \\
  &+& \left.
  (\alpha_V + 3\gamma_S j_\mu j^\mu +\alpha_{TV})\tilde{j}_\nu^\tau\tilde{j}^{\nu\tau}
   +(\delta_V+\delta_{TV})\tilde{j}_\nu^\tau \triangle \tilde{j}^{\nu\tau} \right]\;,
\label{contfield}
\end{eqnarray}
where $\rho_S$ and $j_\mu$ denote the scalar density and baryon
current, respectively , and $\rho_S^\tau$ and $j_\mu^\tau$ are the
corresponding neutron (proton) densities and currents. For protons
there is an additional contribution from the Coulomb interaction:
\begin{equation}
\partial_{\xi}^2\mathcal{E}^{(\xi)}_{C} = e^2\int\int d{\bm r }d{\bm r^\prime }
       \frac{\tilde{\rho}^p(\bm r)\tilde{\rho}^p(\bm r^\prime)}
       {| \bm{r}-\bm{r^\prime}|^2}\;.
\label{contcoul}
\end{equation}
To evaluate the contribution of the pairing energy Eq.~(\ref{epair})
to the second derivative of the energy functional,
one needs the non-hermitian response pairing tensor
\begin{eqnarray}
\tilde{\kappa}^\tau(\bm{r})&=&
      -4\sum_{k>0}f_k^2u_k^3v_k \psi^\dagger_k(\bm{r})\psi_k(\bm{r}) \\
\tilde{\kappa}^{\tau*}(\bm{r})&=&
      4\sum_{k>0}f_k^2u_k v_k^3 \psi^\dagger_k(\bm{r})\psi_k(\bm{r})\;.
\end{eqnarray}
This leads to a rather simple expression
\begin{equation}
\partial_{\xi}^2\mathcal{E}^{(\xi)\tau}_{pair} = \frac{V_q}{2}\int d\bm{r}
      \tilde{\kappa}^\tau\tilde{\kappa}^{\tau*}\;.
\label{contpair}
\end{equation}
Finally, by inserting Eqs. (\ref{contfield}), (\ref{contcoul}), and
(\ref{contpair}) into Eq.~(\ref{lambda2}), one obtains the
Lipkin-Nogami parameter $\lambda_{2,\tau}$.
\bigskip
\noindent
\bigskip \bigskip

\bigskip \bigskip
\leftline{\bf ACKNOWLEDGMENTS} This work has been supported in part
by the Bundesministerium f\"ur Bildung und Forschung - project 06 MT
246, by the Gesellschaft f\"ur Schwerionenforschung GSI - project
TM-RIN, and by the Alexander von Humboldt Stiftung.
\bigskip


\newpage
\newpage
\begin {table}[tbp]
\begin {center}
\caption{Excitation energies, spectroscopic quadrupole moments and
B(E2, J $\to$ J-2) values for the one-phonon ($2_1^+$) state,
and the two-phonon triplet ($0_2^+$, $2_2^+$ and $4_1^+$) in $^{32}$S.
GCM calculations performed with particle number projection are compared
with the results reported in Ref.~\protect\cite{BFH.03},
and with available data.}
\bigskip
\begin {tabular}{c|ccc|ccc|c|ccc}
\hline
\hline
\multicolumn{1}{c|}{state}
&\multicolumn{3}{c|}{$E_x$ (MeV)}
&\multicolumn{3}{c|}{$Q_{spec} (e fm^2)$}
&\multicolumn{1}{c|}{Transition}
&\multicolumn{3}{c}{$BE2 (e^2fm^4)$} \\
\hline
     & This work & Ref.~\cite{BFH.03} & Exp.
     & This work & Ref.~\cite{BFH.03} & Exp. &
     & This work & Ref.~\cite{BFH.03} & Exp. \\
\hline
$0_1^+$    &  0.0   & 0.0   &  0.0   &   -  &  -  &  -  & -
           &  -    &   -  &   -   \\
\hline
$2_1^+$    & 2.81   & 3.22 & 2.23  & 5.8  & 2.3  &  -14.9
         & $2_1^+ \to 0_1^+$ & 94   & 38    & 61 \\
\hline
$0_2^+$    & 6.33   & 6.32 & 3.78  & -    &  -   &  -
         & $0_2^+ \to 2_1^+$ & 37   & 144 & 72 \\
\hline
$2_2^+$    & 6.33   & 7.04 & 4.28  & -3.0 & -0.7 & -
         & $2_2^+ \to 2_1^+$ & 131 & 157 & 54 \\
          &        &       &       &     &     & -
         & $2_2^+ \to 0_1^+$ & 0.134 & 0.02 & 11 \\
          &        &       &       &     &     & -
         & $2_2^+ \to 0_2^+$ & 10.7 & 2.8 & - \\
\hline
$4_1^+$    & 6.61   & 7.35  & 4.46  & -1.2 & 11.7 & -
        & $4_1^+ \to 2_1^+$ & 140.5 & 94 & 72 \\
\hline
\hline
\end{tabular}
\label{TabA}
\end{center}
\end{table}
\begin {table}[tbp]
\begin {center}
\caption{Excitation energies and B(E2, J $\to$ J-2) values
for the two lowest bands in $^{32}$S. The results of GCM calculations
performed without particle number projection are shown in comparison
with those of Ref.~\protect\cite{RER.00}.}
\bigskip
\begin {tabular}{c|cc|c|cc}
\hline
\hline
\multicolumn{1}{c|}{state}
&\multicolumn{2}{c|}{$E_x$ (MeV)}
&\multicolumn{1}{c|}{Transition}
&\multicolumn{2}{c}{$BE2 (e^2fm^4)$} \\
\hline
     & This work & Ref.~\cite{RER.00} &
     & This work & Ref.~\cite{RER.00}  \\
\hline
$0_1^+$   & 0.0  & 0.0  &  -  &  -   & -   \\
$2_1^+$   & 2.181  & 2.107  &  $2_1^+ \to 0_1^+$  & 66.2  & 72.3  \\
$4_1^+$   & 5.395   & 5.825 & $4_1^+ \to 2_1^+$  & 102.9    &  119.8 \\
$6_1^+$   & 9.661   & 10.962 & $6_1^+ \to 4_1^+$  & 146.1    &  142.8 \\
\hline
$0_3^+$   & 3.24  & 3.778  &  -  &  -   & -   \\
$2_2^+$   & 4.832  & 4.282  &  $2_2^+ \to 0_3^+$  & 33.3  & 58.0  \\
$4_2^+$   & 9.213   & 9.097 & $4_2^+ \to 2_2^+$  & 121.1    &  132.2 \\
\hline
\hline
\end{tabular}
\label{TabB}
\end{center}
\end{table}
\begin {table}[tbp]
\begin {center}
\caption{Excitation energies, spectroscopic quadrupole moments, and
B(E2, J $\to$ J-2) values for the ground-state band in
$^{36}$Ar. GCM calculations performed with particle number projection
are compared with those of Ref.~\protect\cite{BFH.03}, and with
available data.}
\bigskip
\begin {tabular}{c|ccc|ccc|c|ccc}
\hline
\hline
\multicolumn{1}{c|}{state}
&\multicolumn{3}{c|}{$E_x$ (MeV)}
&\multicolumn{3}{c|}{$Q_{spec} (e fm^2)$}
&\multicolumn{1}{c|}{Transition}
&\multicolumn{3}{c}{$BE2 (e^2fm^4)$} \\
\hline
     & This work & Ref.~\cite{BFH.03} & Exp.
     & This work & Ref.~\cite{BFH.03} & Exp. &
     & This work & Ref.~\cite{BFH.03} & Exp. \\
\hline
$0_1^+$    &  0.0   & 0.0   &  0.0   &   -  &  -  &  -  & -
           &  -    &   -  &   -   \\
\hline
$2_1^+$    & 2.26   & 2.8 & 1.97  & 13  & 13  &  -
         & $2_1^+ \to 0_1^+$ & 79   & 44    & 60$\pm$ 6 \\
\hline
$4_1^+$    & 6.40   & 7.43 & 4.41  & 14    &  12   &  -
         & $4_1^+ \to 2_1^+$ & 129   & 103 & - \\
\hline
$6_2^+$    & 13.94   & 13.65 & 9.18  & 9.9 & -1.3 & -
         & $6_2^+ \to 4_1^+$ & 152 & 93 & - \\
\hline
\hline
\end{tabular}
\label{TabC}
\end{center}
\end{table}
\begin {table}[tbp]
\begin {center}
\caption{Excitation energies and B(E2, J $\to$ J-2) values
for the ground-state band in $^{36}$Ar. GCM calculations performed
without particle number projection are compared with the results
reported in Ref.~\cite{RER.04}.}
\bigskip
\begin {tabular}{c|cc|c|cc}
\hline
\hline
\multicolumn{1}{c|}{state}
&\multicolumn{2}{c|}{$E_x$ (MeV)}
&\multicolumn{1}{c|}{Transition}
&\multicolumn{2}{c}{$BE2 (e^2fm^4)$} \\
\hline
     & This work & Ref.~\cite{RER.04} &
     & This work & Ref.~\cite{RER.04}  \\
\hline
$0_1^+$   & 0.0  & 0.0  &  -  &  -   & -   \\
$2_1^+$   & 1.54  & 1.45 &  $2_1^+ \to 0_1^+$  & 74.8  & 72.1  \\
$4_1^+$   & 4.99   & 4.54 & $4_1^+ \to 2_1^+$  & 114.7    &  102.3 \\
$6_1^+$   & 12.15   & 10.01 & $6_1^+ \to 4_1^+$  & 142.4    &  112.8 \\
\hline
\hline
\end{tabular}
\label{TabD}
\end{center}
\end{table}
\end{document}